\documentclass[aps]{revtex4-2}

\usepackage{eurosym}
\usepackage{graphicx}
\usepackage[colorlinks=true, pdfstartview=FitV, linkcolor=blue, citecolor=red, urlcolor=magenta]{hyperref}
\usepackage{graphicx}
\usepackage{latexsym}
\usepackage{amsmath}
\usepackage{amsfonts}
\usepackage{amssymb}
\usepackage{verbatim}
\usepackage{braket}
\usepackage{extarrows}
\usepackage{hyperref}
\usepackage{orcidlink}
\usepackage{geometry}
\usepackage[english]{babel} 
\usepackage{tikz}
\usetikzlibrary{arrows.meta, decorations.pathmorphing, 3d,calc}

\begin{document}

\title{Vacuum Energy and Topological Mass from a Constant Magnetic Field and Boundary Conditions in Coupled Scalar Field Theories}

\author{$^{1}$A.~J.~D.~Farias~Junior~\orcidlink{0000-0001-5480-546X}}
\email{antonio.farias@ifal.edu.br}

\author{$^{2}$Andrea~Erdas~\orcidlink{0000-0002-9658-8771}}
\email{aerdas@loyola.edu}

\author{$^{3}$Herondy~F.~Santana~Mota~\orcidlink{0000-0002-7470-1550}}
\email{hmota@fisica.ufpb.br}

\affiliation{$^{1}$Instituto Federal de Alagoas,\\
CEP 57460-000, Piranhas, Alagoas, Brazil}

\affiliation{$^{2}$Department of Physics, Loyola University Maryland,\\
4501 North Charles Street, Baltimore, Maryland 21210, USA}

\affiliation{$^{3}$Departamento de F\'{\i}sica, Universidade Federal da Para\'{\i}ba,\\
Caixa Postal 5008, Jo\~{a}o Pessoa, Para\'{\i}ba, Brazil}

\begin{abstract}

We investigate the combined effects of a uniform magnetic field and boundary conditions on vacuum energy and topological mass generation in a coupled scalar field theory. The system consists of a real scalar field, subject to Dirichlet boundary conditions, interacting via self- and cross-couplings with a gauge-coupled complex scalar field obeying mixed boundary conditions between two perfectly reflecting parallel plates. The magnetic field induces Landau quantization, leading to novel contributions. Employing zeta-function regularization within the effective potential formalism, we derive the renormalized effective potential up to second order in the coupling constants without imposing a vanishing magnetic field in the renormalization scheme. Our renormalization approach preserves magnetic contributions while properly removing divergences, enabling a consistent treatment of finite-size corrections, magnetic effects, and interaction terms. We compute the vacuum energy per unit area of the plates, analyze the emergence of a topological mass from boundary and magnetic contributions, and evaluate the first-order coupling-constant corrections at two-loop order. Detailed asymptotic analysis are presented for both weak- and strong-field regimes, revealing exponential suppression at high magnetic fields and nontrivial polynomial and logarithmic behavior in the weak-field limit.
\end{abstract}

\maketitle

\section{Introduction}
\label{intro}

The study of quantum field theory in confined geometries has long revealed a variety of intriguing phenomena, most notably the Casimir effect \cite{casimir1948attraction}, in which boundary conditions modify the vacuum fluctuations of quantum fields, leading to measurable forces between the confining surfaces. Originally, these surfaces were considered as perfectly reflecting, parallel, and neutral plates, in which the quantum modes of the electromagnetic field were confined \cite{bressi2002measurement, Sparnaay:1958wg} (see also Refs.~\cite{lamoreaux1997demonstration, mohideen1998precision, kim2008anomalies, wei2010results} for other geometries). When additional ingredients, such as external fields or different boundary conditions, are present, the spectral properties of the field modes change in nontrivial ways, giving rise to qualitatively new physical effects. In particular, a uniform magnetic field couples to charged modes and reorganizes their spectrum into Landau levels, thereby reshaping the vacuum energy and the associated forces. Such scenarios are relevant in contexts ranging from strongly magnetized condensed-matter systems to high-energy physics.

While the Casimir-like effect in the presence of magnetic fields has been explored for various field contents and boundary conditions, most previous studies have focused on single-field models or on purely spectral-sum approaches \cite{Cougo-Pinto:1998jun, Elizalde:2002kb, Ostrowski:2005rm, Sitenko:2014wwa, blau1991analytic}. In the context of Lorentz-violation scenarios, the influence of a constant magnetic field has been analyzed in Refs.~\cite{erdas2020casimir, Erdas:2024bxd, Erdas:2025gbv}, and with thermal corrections included in Refs.~\cite{Erdas:2021xvv, Erdas:2013jga, Erdas:2013dha, Erdas:2015yac, Erdas:2025ffm} (see also Refs.~\cite{blau1991analytic, Elizalde:2002kb, Cougo-Pinto:1998jwo, Ostrowski:2005rm, Sitenko:2015eza, Erdas:2010mz, Erdas:2023wzy} for studies involving fermion fields). In addition, the Salam-Weinberg theory of the electroweak interaction in the presence of a constant magnetic field has been considered in Ref.~\cite{Erdas:1990gy}.

In the past several years, increasing attention has been devoted to the analysis of vacuum energy corrections in the presence of boundary conditions, nontrivial topologies, and external fields \cite{bordag2009advances, elizalde1995zeta, milton2001casimir, toms1980interacting, toms1980symmetry, porfirio2021ground, Cruz:2020zkc, Aj, PhysRevD.107.125019}. These scenarios not only modify the zero-point energy but also give rise to topological contributions to the effective mass of the fields \cite{toms1980interacting, toms1980symmetry, PhysRevD.107.125019, Junior:2024smu}, which can play a decisive role in the stability of the vacuum and in the occurrence of symmetry-breaking transitions \cite{Flachi:2024ztd}. The so-called topological mass arises from the combined effect of quantum fluctuations and the finite-size geometry, and its sign can be altered by varying the strength of the external field or the coupling constants of the theory.

The investigation of quantum field theories under external constraints such as boundary conditions and
background fields represents a fundamental area of research with wide-ranging implications across theoretical physics~\cite{Milton:2010zz, BirrellDavies1982, AltlandSimons2010}. This work examines the combined effects of a constant magnetic field and boundary conditions on vacuum energy and topological mass generation in a system of coupled scalar fields.

Specifically, our study focuses on a physical system consisting of a real scalar field $\psi$ interacting with a complex scalar field $\phi$ through both self-couplings and cross-couplings, in the presence of a constant magnetic field $B$ oriented perpendicularly to two perfectly reflecting parallel plates. The real field satisfies Dirichlet boundary conditions at the plates, while the gauge-coupled complex field obeys mixed boundary conditions. The novel aspects of our work include a comprehensive treatment of real and complex scalar fields with distinct boundary conditions under a uniform magnetic field; the derivation of the renormalized effective potential up to first order in the coupling constants; a precise calculation of the vacuum energy with finite-size corrections; a detailed analysis of topological mass generation through boundary effects and magnetic interactions; and a systematic examination of both weak- and strong-field regimes.
 
The methodological approach combines zeta-function regularization technique with effective potential formalism to handle divergences while maintaining physical transparency throughout the calculations. Our renormalization procedure carefully preserves magnetic field contributions, distinguishing this work from alternative treatments that impose $B=0$ as a renormalization condition. This distinction proves crucial for obtaining physically meaningful results in confined geometries with persistent external fields.

The paper is organized as follows. Section~\ref{secII} introduces the model and outlines the effective potential formalism in the presence of a uniform magnetic field and boundary conditions. Section~\ref{secIII} presents the calculation of the generalized zeta functions and the derivation of the renormalized effective potential up to first order in the coupling constants. In Sec.~\ref{secIV}, we analyze the vacuum energy, the topological mass, and the first-order coupling corrections to the vacuum energy, along with a detailed asymptotic analysis for weak and strong magnetic field. Finally, our conclusions and perspectives for future developments are summarized in Sec.~\ref{secV}.

Through this paper we use natural units in which both the Planck constant
and the speed of light are set equal to one, $\hslash =c=1$. The Lorentzian signature of the metric to be adopted here is (+, -, -, -).

\section{Coupled Scalar Field Theories and the Effective Potential}
\label{secII}
\subsection{Complex Scalar Field Action and Gauge Coupling}
\label{secIIA}
In this section, we present the coupled scalar field theories that will be considered in our study. We begin with the Lagrangian for a complex scalar field $\phi$ coupled to a gauge field $A_\mu$, given by \cite{Schwartz:2014sze, ryder1996quantum, greiner2013field}:
\begin{equation}
\mathcal{L} =
\left( \partial^{\mu} \phi^{\ast} - i e A^{\mu} \phi^{\ast} \right)
\left( \partial_{\mu} \phi + i e A_{\mu} \phi \right)
- m^{2} \phi^{\ast} \phi \,,
\end{equation}
where $e$ is the coupling constant associated with the electric charge of the complex field, and $m$ is its mass.

By separating the real and imaginary parts of the complex field, the above Lagrangian can be decomposed into three parts, namely,
\begin{eqnarray}
\mathcal{L} &=& \mathcal{L}_{1} + \mathcal{L}_{2} + \mathcal{L}^{\prime },
\label{threeP}
\end{eqnarray}
where
\begin{eqnarray}
\mathcal{L}_{j} &=& \frac{1}{2}\partial^{\mu} \varphi_{j} \, \partial_{\mu} \varphi_{j}
+ \frac{e^{2} A^{\mu} A_{\mu}}{2} \, \varphi_{j}^{2}
- \frac{1}{2} m^{2} \varphi_{j}^{2} \,,
\notag \\
\mathcal{L}^{\prime} &=& e A^{\mu} \left( \varphi_{1} \, \partial_{\mu} \varphi_{2}
- \varphi_{2} \, \partial_{\mu} \varphi_{1} \right) \,,
\label{g8}
\end{eqnarray}
with $\varphi_j$ $(j = 1, 2)$ denoting the real components of the complex field. 
Note that the last term in Eq.~\eqref{threeP} does not contribute to the effective potential, since the mixed derivatives vanish when expanding the action.

Let us now consider the action for a given component $j$ of the complex field by using $\mathcal{L}_{j}$ in Eq.~\eqref{g8}. Thus, we can write 
\begin{equation}
S_{j} \left[ \varphi_{j} \right] = \frac{1}{2} \int d^{4}x \,
 \left( \partial^\mu\varphi_j\partial_\mu\varphi_j+{e^2A^\mu A_\mu}\varphi_j^2-m^2\varphi_j^2 \right)\,.
\label{g11}
\end{equation}

Next, we will carefully add a total divergence term. Before doing that, being aware of the fact that in the presence of boundaries a total divergence term produces a contribution dependent on the value that the fields or their derivatives take on the boundaries, we specify the mixed boundary conditions satisfied by the field $\varphi_j$
\begin{eqnarray}
\varphi_j(t,x,y,L) &=&0 \,,
\notag \\
\partial_z\varphi_j(t,x,y,0) &=&0 \,,
\label{g8a}
\end{eqnarray}
where $\partial_z\equiv {\partial\over\partial z}$. Now we add the following total divergence to the integrand of the action
\begin{equation}
- \frac{1}{2}  \partial_\mu\left[\varphi_j(\partial^\mu +ieA^\mu)\varphi_j\right]
\end{equation}
which, according to the mixed boundary conditions of Eq.~\eqref{g8a} and to $\partial_\mu A^\mu=0$, the gauge invariance condition of $A^\mu$, vanishes at the boundaries, and obtain
\begin{equation}
S_{j} \left[ \varphi_{j} \right] = \frac{1}{2} \int d^{4}x \,
\varphi_{j} \left( 
- \partial^{\mu} \partial_{\mu}
- 2 e A^{\mu} i \partial_{\mu}
+ e^{2} A^{\mu} A_{\mu}
- m^{2} \right) \varphi_{j} \,.
\label{g11a}
\end{equation}

Next, we consider a gauge field given by
\begin{equation}
A^{\mu} = \left( 0, 0, Bx, 0 \right),
\label{GField}
\end{equation}
which generates a constant magnetic field $B$ along the $z$-direction.  
Furthermore, applying both the Wick rotation $t = -i\tau$ and Eq.~\eqref{GField} to Eq.~\eqref{g11} allows us to express the action in Euclidean spacetime with coordinates $(\tau, x, y, z)$. Explicitly, we obtain
\begin{eqnarray}
S_{j} \left[ \varphi_{j} \right] 
&=& -\frac{1}{2} \int d^{4}x \, \varphi_{j} \left[
 -\partial_{\tau}^{2}
 -\partial_{x}^{2}
 -\partial_{z}^{2}
 + \left( i \partial_{y} - e B x \right)^{2}
 + m^{2} \right] \varphi_{j} 
\notag \\
&=& \frac{1}{2} \int d^{4}x \, \varphi_{j} \, \square_{B} \, \varphi_{j},
\label{E_action}
\end{eqnarray}
where the operator $\square_{B}$ is defined as
\begin{equation}
\square_{B} = \partial_{\tau}^{2} + \partial_{x}^{2} + \partial_{z}^{2} 
- \left( i \partial_{y} - e B x \right)^{2} \,.
\label{Box_B}
\end{equation}

In the next subsection, we will write the total action of the coupled scalar field theories under consideration and discuss how to obtain the expression for the effective potential up to second order.

\subsection{Total Action and Effective Potential}
\label{secIIB}
Here, we consider a real scalar field $\psi$ coupled to a charged complex scalar field whose components are $\varphi_j$ $(j = 1, 2)$, as shown previously. In addition, we include self-coupling contributions for both fields. In this case, in Euclidean coordinates, the total action can be written as \cite{PhysRevD.107.125019}
\begin{eqnarray}
S_{E} \left[ \psi, \varphi_{j} \right] 
&=& \frac{1}{2} \int d^{4}x 
\left[ \psi \, \square \, \psi 
+ \sum_{j=1}^{2} \varphi_{j} \, \square_{B} \, \varphi_{j} \right]
- \int d^{4}x \, U \left( \psi, \varphi \right),
\label{rc2}
\end{eqnarray}
where 
\begin{eqnarray}
U \left( \psi, \varphi \right) 
&=& \frac{m^{2} + C_{2}}{2} \, \psi^{2}
+ \frac{\mu^{2}}{2} \, \varphi^{2}
+ \frac{g}{2} \, \varphi^{2} \psi^{2}
+ \frac{\lambda_{\varphi}}{4!} \, \varphi^{4}
+ \frac{\lambda_{\psi} + C_{1}}{4!} \, \psi^{4}
+ C_{3} \,.
\label{rc21}
\end{eqnarray}
In the above expressions, $m$ is the mass of the real field $\psi$, $\mu$ is the mass of the complex field $\phi$, $\lambda_{\psi}$ and $\lambda_{\varphi}$ are the self-coupling constants for the real and complex fields, respectively, and $g$ is the coupling constant associated with the interaction between the scalar fields. Here, $\varphi^{2} = \varphi_{1}^{2} + \varphi_{2}^{2}$, and $\square$ denotes the usual d’Alembertian operator, which in Euclidean coordinates is expressed as
\begin{equation}
\square = \partial_{\tau}^{2} + \nabla^{2} \,.
\end{equation}
Note that the operator $\square_{B}$ has been defined in Eq.~\eqref{Box_B}, and $C_{i}$ $(i = 1, 2, 3)$ are the renormalization constants, whose explicit form will be determined in the renormalization process of the effective potential, presented in the next section.

The development of the effective potential approach has been presented in several works (see, for example, Refs.~\cite{toms1980symmetry, Aj, PhysRevD.107.125019}), so here we only summarize the most relevant aspects. In this framework, the expansion of the effective potential up to second order can be written as
\begin{equation}
V_{\mathrm{eff}} \left( \Psi \right)
= V^{(0)} \left( \Psi \right)
+ V^{(1)} \left( \Psi \right)
+ V^{(2)} \left( \Psi \right) .
\label{rc2.1}
\end{equation}
The zeroth-order term, $V^{(0)}(\Psi)$, corresponds to the classical potential, i.e., the tree-level contribution, and is given by
\begin{eqnarray}
V^{(0)} \left( \Psi \right) &=& U \left( \Psi \right) \notag \\
&=& \frac{m^{2} + C_{2}}{2} \, \Psi^{2}
+ \frac{\lambda_{\psi} + C_{1}}{4!} \, \Psi^{4}
+ C_{3} .
\label{rc2.2}
\end{eqnarray}
Here, $\Psi$ is the constant background field about which the real scalar field is expanded, as is standard in the path-integral quantization procedure. In contrast, the constant background field associated with the complex field has been set to zero. This choice is justified because we impose mixed boundary conditions on the complex field, and the only constant field configuration that satisfies such conditions is the trivial one, $\Phi = 0$ \cite{toms1980symmetry, PhysRevD.107.125019}.

The effective potential can also be represented in terms of 1PI (one-particle irreducible) vacuum bubble Feynman diagrams as follows:
\begin{eqnarray}
V_{\mathrm{eff}}(\Psi) = 
\underbrace{
\begin{tikzpicture}[baseline=-0.5ex]
  \draw[thick] (-0.6,0) -- (0.6,0);
\end{tikzpicture}
}_{\text{tree level}}
+ \
\underbrace{
\begin{tikzpicture}[baseline=-0.5ex]
  \draw[thick] (0,0) circle (0.4);
\end{tikzpicture}
}_{\text{1-loop}}
+ \
\underbrace{
\begin{tikzpicture}[baseline=-0.5ex]
  \draw[thick] (0,0) circle (0.4);
  \draw[thick] (0.8,0) circle (0.4);
  \fill (0.4,0) circle (2pt); 
\end{tikzpicture}
}_{\text{2-loop}}
+ \cdots\,,
\label{Fdiagrams0}
\end{eqnarray}
where the first- and second-order contributions are illustrated explicitly.  
The two-loop vacuum contribution to the effective potential will be analyzed later using the appropriate Feynman rules in Sec.~\ref{secIVC}.

In particular, the one-loop contribution to the effective potential can be expressed in terms of generalized zeta functions \cite{toms1980symmetry, Aj, PhysRevD.107.125019}, which arise from the following functional integral:
\begin{equation}
V^{(1)}(\Psi) = - \frac{1}{\Omega_4} \ln \int \mathcal{D}\psi \, \mathcal{D}\varphi_1 \, \mathcal{D}\varphi_2 \, \exp \left[ -\frac{1}{2} \left( \psi, \hat{A} \psi \right) - \frac{1}{2} \left( \varphi_1, \hat{B} \varphi_1 \right) - \frac{1}{2} \left( \varphi_2, \hat{B} \varphi_2 \right) \right],
\label{rc2.3}
\end{equation}
where $\Omega_4$ is the four-dimensional volume of Euclidean spacetime, which depends on the boundary conditions imposed on the fields. We have also used the notation
\begin{equation}
\left( \psi, \hat{A} \psi \right) = \int d^{4}x \, \psi(x) \, \hat{A} \, \psi(x),
\label{rc2.4}
\end{equation}
with the self-adjoint differential operators $\hat{A}$ and $\hat{B}$ defined as
\begin{equation}
\hat{A} = -\square + m^{2} + \frac{\lambda_{\psi}}{2} \Psi^{2}, \quad \quad \hat{B} = -\square_B + \mu^{2} + g \Psi^{2}.
\label{rc2.5}
\end{equation}

By evaluating the path integral in Eq.~\eqref{rc2.3}, one can show that the one-loop correction to the classical potential $U(\Psi)$ separates into two contributions \cite{PhysRevD.107.125019}:
\begin{equation}
V^{(1)}(\Psi) = V_{\psi}^{(1)}(\Psi) + V_{\varphi}^{(1)}(\Psi),
\label{rc14}
\end{equation}
where
\begin{equation}
V_{\psi}^{(1)} = -\frac{1}{2 \Omega_4} \left[ \zeta_{\psi}'(0) + \zeta_{\psi}(0) \ln \alpha^{2} \right],
\label{rc141}
\end{equation}
is the contribution from the real scalar field, and
\begin{equation}
V_{\varphi}^{(1)} = -\frac{1}{\Omega_4} \left[ \zeta_{\varphi}'(0) + \zeta_{\varphi}(0) \ln \beta^{2} \right],
\label{rc142}
\end{equation}
is the contribution from the complex scalar field.  
The mass-dimensional constants $\alpha$ and $\beta$ serve as integration measure in the functional spaces of the real and complex fields, respectively, and will be removed through the renormalization of the effective potential.  

The generalized zeta function $\zeta(s)$ is defined as
\begin{equation}
\zeta(s) = \sum_{\sigma} \Lambda_{\sigma}^{-s},
\label{zeta}
\end{equation}
where $\sigma$ represents the set of quantum numbers of the system and $\Lambda_{\sigma}$ denotes the eigenvalues of a self-adjoint differential operator $\hat{\mathcal{O}}$, such as those given in Eq.~\eqref{rc2.5}. The series converges for $\mathrm{Re}(s) > 2$ and is regular at $s=0$, admitting an analytic continuation to other values of $s$ \cite{elizalde1995zeta, elizalde1995ten}.

Therefore, to obtain the one-loop contribution in Eq.~\eqref{rc14} to the effective potential, we need to solve the eigenvalue problem
\begin{equation}
\hat{\mathcal{O}} \eta = \Lambda \eta,
\label{EVP}
\end{equation}
where $\eta$ represents quantum fluctuations of either $\psi$ or $\varphi_j$ around their respective background fields $\Psi$ and $\Phi=0$.

Note that to renormalize the effective potential, standard renormalization conditions will be employed, which will be explicitly presented in the next section.
\section{Real and Complex Scalar Fields Subject to Boundary Conditions}
\label{secIII}
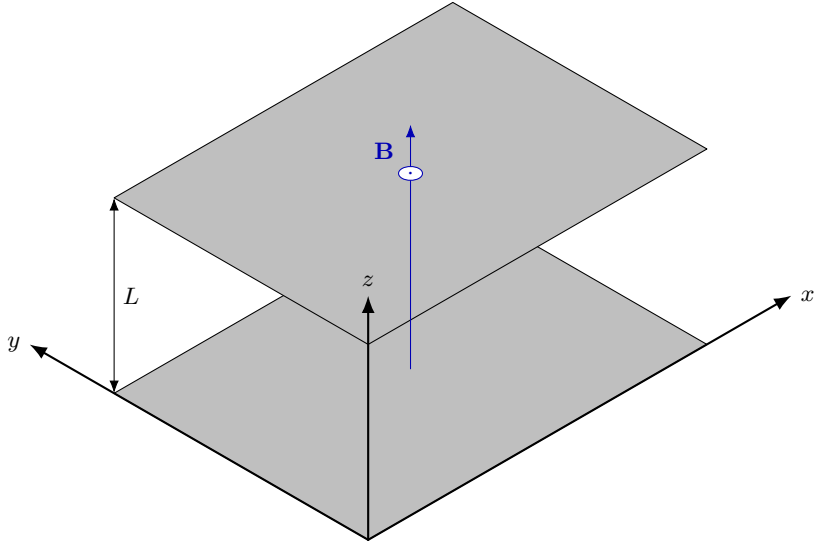
\begin{figure}[h]
\centering
\begin{tikzpicture}[scale=1.3, line join=round, line cap=round, >=Latex,
                    x={(0.866cm,0.5cm)}, y={( -0.866cm,0.5cm)}, z={(0cm,1cm)}]
  \def\Lx{4}   
  \def\Ly{3}   
  \def\zsep{2} 

  \fill[gray!50] (0,0,0) -- (\Lx,0,0) -- (\Lx,\Ly,0) -- (0,\Ly,0) -- cycle;
  \draw[black] (0,0,0) -- (\Lx,0,0) -- (\Lx,\Ly,0) -- (0,\Ly,0) -- cycle;

  \fill[gray!50] (0,0,\zsep) -- (\Lx,0,\zsep) -- (\Lx,\Ly,\zsep) -- (0,\Ly,\zsep) -- cycle;
  \draw[black] (0,0,\zsep) -- (\Lx,0,\zsep) -- (\Lx,\Ly,\zsep) -- (0,\Ly,\zsep) -- cycle;

\draw[->, thin, blue!70!black] (\Lx/2, \Ly/2, 0) -- (\Lx/2, \Ly/2, \zsep+0.5);

\draw[blue!70!black, fill=white] (\Lx/2, \Ly/2, \zsep) circle (0.1) node {$\cdot$};

\node[blue!70!black, right] at (\Lx/2 + 0.97, \Ly/1, \zsep/2) {$\mathbf{B}$};

  \draw[->,thick] (0,0,0) -- (5,0,0) node[right] {$x$};
  \draw[->,thick] (0,0,0) -- (0,4,0) node[left] {$y$};
  \draw[->,thick] (0,0,0) -- (0,0,2.5) node[above] {$z$};

\draw[<->, thin, black] 
    (0, \Ly, 0) -- node[right] {$L$} (0, \Ly, \zsep);

\end{tikzpicture}
\caption{Schematic view of two perfectly reflecting parallel plates in the $x$-$y$ plane separated by $L$ along the $z$-axis, with a uniform magnetic field $\mathbf{B}$ directed along $z$. The real scalar field $\psi$ satisfies Dirichlet boundary conditions, while the gauge-coupled complex scalar field $\phi$ obeys mixed boundary conditions.}
\label{fig1}
\end{figure}

In this section, we calculate the generalized zeta function associated with real and complex scalar fields subject to Dirichlet and mixed boundary conditions, respectively. To this end, we consider two perfectly reflecting, parallel plates perpendicular to the $z$-direction, with $L$ being the separation between them. One plate is located at $z=0$ and the other at $z=L$, so that the boundary conditions are applied exactly on the plates. Moreover, we also consider a constant magnetic field in the $z$-direction, whose associated gauge potential is coupled to the complex scalar field (see Fig.~\ref{fig1}), as discussed in the previous section. As a consequence of imposing these boundary conditions on the scalar fields, we also obtain the renormalized effective potential up to first order in each case.

As stated previously, we have to solve the eigenvalue problem of Eq.~\eqref{EVP}, with the operators defined in Eq.~\eqref{rc2.5}. For the quantum fluctuation of the real scalar field, $\eta_{\psi}$, associated with the operator $\hat{A}$, we impose Dirichlet boundary conditions at the plates, i.e.,
\begin{equation}
\eta_{\psi}\left( t,x,y,0\right) = \eta_{\psi}\left( t,x,y,L\right) =0\,.
\label{l1}
\end{equation}%
The explicit form of the eigenfunctions of $\hat{A}$ is not essential here, as it is well known in the literature \cite{bordag2009advances, mostepanenko1997casimir}. The relevant quantity for our purposes is the spectrum of eigenvalues,
\begin{eqnarray}
\Lambda _{\sigma } &=&k_{\tau }^{2}+k_{x}^{2}+k_{y}^{2}+\left( \frac{n\pi }{%
L}\right) ^{2}+M_{\lambda }^{2},\quad n=1,2,\ldots, \notag \\
M_{\lambda }^{2} &=&m^{2}+\frac{\lambda _{\psi }}{2}\Psi ^{2},  
\label{rc14.1}
\end{eqnarray}%
where $(k_{\tau}, k_x, k_y)$ are the continuous momentum components and $n$ is the discrete one. The set of quantum numbers for the real scalar field is thus $\sigma=(k_{\tau}, k_x, k_y, n)$.

For the quantum fluctuations of the complex scalar field, $\eta_j$ $(j=1,2)$, we impose mixed boundary conditions as in Eq.~\eqref{g8a}:
\begin{eqnarray}
\eta _{j}\left( \tau ,x,y,L\right) &=& 0, \notag\\
\partial_z\eta _{j}\left( \tau ,x,y,0\right) &=& 0,
\end{eqnarray}%
where $\partial_z\equiv\frac{\partial}{\partial z}$. In this case, the eigenvalue problem \eqref{EVP} involves the operator $\hat{B}$ in Eq.~\eqref{rc2.5}. The coupling of the gauge field \eqref{GField} to the complex scalar field leads to Landau quantization, and as a consequence the spectrum of $\hat{B}$ is given by
\begin{eqnarray}
\Lambda_{\sigma } &=&k_{\tau }^{2}+eB\left( 2n+1\right) +\left( j+\frac{1}{2}%
\right) ^{2}\frac{\pi ^{2}}{L^{2}}+M_{g}^{2},  \notag \\
M_{g}^{2} &=&\mu ^{2}+g\Psi ^{2},\quad n=0,1,2,\ldots,\quad j=0,1,2,\ldots\,.
\label{rc16.1}
\end{eqnarray}
Here, the quantum numbers are $\sigma = (k_{\tau}, n, j)$, consisting of two discrete indices $(n, j)$ and one continuous momentum component $k_{\tau}$. For illustration, Fig.~\ref{fig1} depicts the two parallel plates in the presence of a constant magnetic field $B$ oriented along the $z$-axis.

\subsection{Generalized Zeta Functions and Non-Renormalized Effective Potential}
\label{secIIIA}
Let us now calculate the generalized zeta function associated with the real scalar field. To do so, we use the definition in Eq.~\eqref{zeta} together with the eigenvalues in Eq.~\eqref{rc14.1}. The summation over $\sigma$ implies that we must integrate over the continuous momenta and sum over the discrete one. This leads to
\begin{equation}
\zeta_{\psi}(s) = \frac{\Omega_{3}}{(2\pi)^{3}} \int dk_{\tau} dk_{x} dk_{y} \sum_{n=1}^{\infty} \left[ k_{\tau}^{2} + k_{x}^{2} + k_{y}^{2} + \left( \frac{n\pi}{L} \right)^{2} + M_{\lambda}^{2} \right]^{-s},
\end{equation}
where $\Omega_{3} = L_{\tau} L_{y} L_{x}$ is a volume-like parameter associated with the continuous momenta.

Subsequently, it is convenient to rewrite the above expression for the zeta function in a form that allows us to perform the momentum integrals explicitly. This is achieved by using the following identity:
\begin{equation}
w^{-s} = \frac{2}{\Gamma(s)} \int_0^\infty d\rho \, \rho^{2s-1} e^{-w \rho^2}\,,
\label{i1}
\end{equation}
where $\Gamma(s)$ is the gamma function. Notice that the integrals over the momenta that appear after using this representation are Gaussian-like and can be straightforwardly evaluated. The remaining integral in $\rho$ can then be identified with the integral representation of the gamma function \cite{abramowitz1965handbook},
\begin{equation}
\Gamma(s) = \int_0^\infty dt \, t^{s-1} e^{-t}\,.
\label{i2}
\end{equation}
Applying these steps, the generalized zeta function associated with the real scalar field can be written as
\begin{equation}
\zeta_{\psi}(s) = \frac{\Omega_3 \pi^{\frac{3}{2} - 2s}}{8 L^{3 - 2s}} \frac{\Gamma\left(s - \frac{3}{2}\right)}{\Gamma(s)} \sum_{n=1}^\infty \left[ n^2 + \frac{M_\lambda^2 L^2}{\pi^2} \right]^{\frac{3}{2} - s}.
\label{zf1}
\end{equation}

To perform the sum over the discrete quantum number $n$, we employ the Epstein-Hurwitz zeta function representation \cite{Elizalde:1995hck, Elizalde_hiroshima}, which reads
\begin{align}
\sum_{l=1}^\infty \left[l^2 + \kappa^2 \right]^{\frac{3}{2} - s} 
= & -\frac{\kappa^{3 - 2s}}{2} + \frac{\sqrt{\pi}}{2} \frac{\Gamma(s - 2)}{\Gamma\left(s - \frac{3}{2}\right)} \kappa^{4 - 2s} \notag \\
& + \frac{2 \kappa^{2 - s} \pi^{s - \frac{3}{2}}}{\Gamma\left(s - \frac{3}{2}\right)} \sum_{j=1}^\infty j^{s-2} K_{s-2}\left(2 \pi j \kappa \right),
\end{align}
where $K_\nu(z)$ is the modified Bessel function of the second kind. Using this, the generalized zeta function in Eq.~\eqref{zf1} can be expressed as a sum of three terms,
\begin{align}
\zeta_{\psi}(s) = & - \frac{\Omega_4 M_\lambda^{3 - 2s}}{16 \pi^{\frac{3}{2}} L} \frac{\Gamma\left(s - \frac{3}{2}\right)}{\Gamma(s)} + \frac{\Omega_4 M_\lambda^{4 - 2s}}{16 \pi^2} \frac{\Gamma(s - 2)}{\Gamma(s)} \notag \\
& + \frac{\Omega_4 M_\lambda^{2 - s}}{4 \pi^2 L^{2 - s} \Gamma(s)} \sum_{n=1}^\infty n^{s-2} K_{s-2}\left(2 n M_\lambda L \right),
\label{zeta_D}
\end{align}
where $\Omega_4 = \Omega_3 L$ is the four-dimensional volume of the Euclidean spacetime. By evaluating the zeta function and its derivative in the limit $s \to 0$, we obtain, according to Eq.~\eqref{rc141}, the one-loop correction to the effective potential,
\begin{align}
V_{\psi}^{(1)} = & - \frac{M_\lambda^4}{64 \pi^2} \ln\left(\frac{\alpha^2}{M^2_\lambda}\right) - \frac{3 M_\lambda^4} {128 \pi^2} + \frac{M_\lambda^3}{24 \pi L} \notag \\
& - \frac{M_\lambda^2}{8 \pi^2 L^2} \sum_{n=1}^\infty n^{-2} K_2\left( 2 n M_\lambda L \right).
\label{vd}
\end{align}
Here, the terms independent of $L$ correspond to the usual vacuum contributions, while the terms depending explicitly on the plate separation $L$ describe finite-size corrections due to the boundary conditions. The last term, involving the Bessel functions, encodes the effect of the plates and decays exponentially for large values of $M_\lambda L$. Note that this one-loop effective potential correction has been previously studied in the context of non-interacting massless scalar fields subject to Dirichlet boundary conditions in Ref.~\cite{toms1980symmetry}, and in Ref.~\cite{Cruz:2020zkc} the investigation considered the massive case. In our interacting theory, the result in Eq.~\eqref{vd} provides the basis for including boundary effects into the effective potential analysis.

For the gauge-coupled complex scalar field, the generalized zeta function can be obtained by substituting Eq.~\eqref{rc16.1} into the definition in Eq.~\eqref{zeta}. In this case, the set of quantum numbers $\sigma$ contains one continuous momentum, $k_{\tau}$, and two discrete quantum numbers, $j$ and $n$. This leads to
\begin{equation}
\zeta_{\varphi}(s) = \frac{L_{\tau}}{2\pi} \left( \frac{eB\, L_{x}L_{y}}{2\pi} \right) 
\int dk_{\tau} \sum_{j=0}^{\infty} \sum_{n=0}^{\infty} 
\left[ k_{\tau}^{2} + eB(2n+1) + \left(j+\frac{1}{2}\right)^{2} \frac{\pi^{2}}{L^{2}} + M_{g}^{2} \right]^{-s},
\label{zeta_initial}
\end{equation}
where the factor $\frac{eB\, L_{x}L_{y}}{2\pi}$ accounts for the degeneracy of the Landau levels.  

By making use of the identity in Eq.~\eqref{i1}, the Gaussian-like integral over $k_{\tau}$ can be performed straightforwardly, reducing the problem to the evaluation of the sums over $n$ and $j$. The exponential factors that appear in the integrand have the form  
\begin{equation}
e^{-eB(2n+1)\tau^{2} - \left(j+\frac{1}{2}\right)^{2} \frac{\pi^{2}}{L^{2}} \tau^{2}}.
\label{exponentials}
\end{equation}
The sum over the Landau-level index $n$ can be carried out using the identity \cite{erdas2020casimir}
\begin{equation}
\frac{1}{2\sinh(T)} = \sum_{n=0}^{\infty} e^{-(2n+1)T}, \qquad T > 0,
\label{csch_expansion}
\end{equation}
which rewrites the infinite series of exponentials in terms of a hyperbolic function. This step greatly simplifies the structure of the zeta function, allowing for further analytical treatment of the remaining sum over $j$.

In order to express the sum over $j$ in an alternative form, let us first recall the delta function property, $\int dz\, \delta(z-z_0) f(z) = f(z_0)$ \cite{arfken2011mathematical}, together with the relation
\begin{equation}
\sum_{j=0}^{+\infty} e^{-\left( j+\frac{1}{2}\right) ^{2}\frac{\pi ^{2}}{L^{2}}\tau^2}=\frac{1}{2}\sum_{j=-\infty}^{+\infty} e^{-\left( j+\frac{1}{2}\right) ^{2}\frac{\pi ^{2}}{L^{2}}\tau^2},
\label{sum_j1}
\end{equation}
and the Fourier series identity
\begin{equation}
2\pi \sum_{n=-\infty}^{\infty} \delta\!\left( \theta - 2\pi n \right) 
= \sum_{n=-\infty}^{\infty} e^{i n\theta}.
\label{delta_series}
\end{equation}

By making use of these previous expressions, we can show that
\begin{align}
\frac{1}{2}\sum_{j=-\infty}^{+\infty} e^{-\left( j+\frac{1}{2}\right) ^{2}\frac{\pi ^{2}}{L^{2}}\tau^2} 
&= \frac{L}{2\sqrt{\pi}\,\tau} 
+ \frac{L}{\sqrt{\pi}\,\tau} \sum_{j=1}^{\infty} (-1)^j e^{-\frac{L^2 j^2}{\tau^2}}.
\label{j_identity}
\end{align}

Consequently, substituting Eqs.~\eqref{i1}, \eqref{csch_expansion}, and \eqref{j_identity} into the expression for the zeta function, we obtain  
\begin{equation}
\zeta_{\varphi}(s) 
= \frac{eB\Omega_4}{(2\pi)^2}
\frac{1}{\Gamma(s)} \int_{0}^{\infty} \frac{d\tau\, \tau^{2s - 3}}{\sinh(eB\tau^2)}\, e^{-M_g^2\tau^2}
\left[ \frac{1}{2} + \sum_{j=1}^{\infty} (-1)^j e^{-\frac{j^2 L^2}{\tau^2}} \right],
\label{zeta_1}
\end{equation}
where $\Omega_4=L_{\tau}L_xL_yL$ is the four-dimensional volume of the Euclidean spacetime. This representation is convenient for further analytical continuation in $s$ and for isolating the boundary-dependent contributions to the effective potential.

Note that it is not possible to carry out the $\tau$-integral in the second term of Eq.~\eqref{zeta_1}.  
In contrast, the integral in the first term can be evaluated analytically, yielding the Minkowski-space contribution in the presence of the magnetic field.  
Collecting both contributions, the generalized zeta function associated with the coupled complex scalar field can be written as  
\begin{eqnarray}
\zeta _{\varphi }\left( s\right) &=&
\frac{\Omega _{4}}{16\pi^2}
\frac{\Gamma \left( s-1\right)}{\left( 2eB\right) ^{\,s-2}\Gamma \left( s\right)}\,
\zeta _{H}\left( s-1,c\right)\notag\\
&+& \frac{\Omega _{4}}{4\pi^2 L^{4-2s}}\,\frac{1}{\Gamma(s)}
\sum_{j=1}^{\infty}(-1)^j\,\mathcal{I}_s(j, L, \Psi),
\label{zetaF_varphi}
\end{eqnarray}
where $\zeta_{H}(s,z)$ denotes the Hurwitz zeta function~\cite{elizalde1995zeta}, and after performing the change of variables $\tau \to \tau L$, we define
\begin{eqnarray}
\mathcal{I}_s(j, L, \Psi) 
&=& B_L^2 \int_{0}^{\infty} 
\frac{d\tau\,\tau^{2s -3}}{\sinh(B_L^2\tau^2)} 
e^{-M_g^2L^2\tau^2 - \frac{j^2}{\tau^2}},
\label{integral_B}
\end{eqnarray}
with 
\begin{equation}
B_L=\sqrt{eB}\,L,
\qquad
c=\frac{1}{2}+\frac{M_{g}^{2}}{2eB}.
\label{c_coef}
\end{equation}

Evaluating the generalized zeta function \eqref{zetaF_varphi} and its derivative at $s=0$, we obtain
\begin{equation}
\zeta _{\varphi }(0) 
= -\frac{\Omega _{4}}{4\pi^2}(eB)^2\,\zeta _{H}(-1,c),
\end{equation}
and
\begin{eqnarray}
\zeta _{\varphi }^{\prime }(0) =
&-& \frac{\Omega_{4}\,(eB)^{2}}{4\pi ^{2}}
\left[ \frac{d}{ds}\zeta _{H}\left( s-1,c\right) \right]_{s=0}\notag\\ 
&+& \frac{\Omega_{4}\,(eB)^{2}}{4\pi ^{2}}
\big(\ln 2eB - 1\big)\,\zeta _{H}(-1,c)
 \notag\\
&+& \frac{\Omega _{4}}{4\pi^2 L^4}
\sum_{j=1}^{\infty}(-1)^j\,\mathcal{I}_0(j, L, \Psi),
\end{eqnarray}
where $\mathcal{I}_0(j, L, \Psi)$ is given by Eq.~\eqref{integral_B} evaluated at $s=0$.

From these results, using Eq.~\eqref{rc142}, the one-loop correction to the effective potential reads
\begin{eqnarray}
V_{\varphi }^{(1)} =
&-&\frac{(eB)^{2}}{4\pi^{2}}\,\zeta _{H}(-1,c)\,
\ln \!\left( \frac{2eB}{\beta ^{2}} \right) 
+ \frac{(eB)^{2}}{4\pi^{2}}\,\zeta _{H}(-1,c) \notag\\
&+& \frac{(eB)^{2}}{4\pi^{2}}
\left[ \frac{d}{ds}\zeta _{H}\left( s-1,c\right) \right]_{s=0}  \notag \\
&-& \frac{1}{4\pi^2 L^4} \sum_{j=1}^{\infty}(-1)^j\,\mathcal{I}_0(j, L, \Psi) \,.
\label{vc}
\end{eqnarray}

Finally, combining Eqs.~\eqref{vd} and \eqref{vc}, we obtain the total non-renormalized effective potential up to one-loop order as
\begin{eqnarray}
V\left( \Psi \right) &=&
\frac{m^{2}+C_{2}}{2}\,\Psi ^{2}
+ \frac{\lambda _{\psi} + C_{1}}{4!}\,\Psi ^{4}
+ C_{3}
+ \frac{M_{\lambda }^{4}}{64\pi ^{2}}
\left[ \ln\left( \frac{M^2_{\lambda }}{\alpha^2}\right) - \frac{3}{2}\right]
+ \frac{M_{\lambda }^{3}}{24\pi L} \notag \\
&& + \frac{(eB)^{2}}{4\pi^{2}}\,\zeta _{H}(-1,c) 
\left[ 1 - \ln\left( \frac{2eB}{\beta^{2}}\right)\right] 
+ \frac{(eB)^{2}}{4\pi^{2}} 
\left[ \frac{d}{ds}\zeta _{H}\left( s-1,c\right) \right]_{s=0}  \notag \\
&& - \frac{M_{\lambda }^{2}}{8\pi ^{2}L^{2}} 
\sum_{n=1}^{\infty} n^{-2} K_{2}\left( 2nM_{\lambda }L\right) 
- \frac{1}{4\pi^2 L^4} \sum_{j=1}^{\infty}(-1)^j\,\mathcal{I}_0(j, L, \Psi) \,.
\label{vj1}
\end{eqnarray}

In the next subsection, we proceed with the renormalization of the effective potential, absorbing all divergent contributions into the appropriate counterterms.

\subsection{Renormalization}
Here we consider the effective potential as given by Eq.~\eqref{vj1}.  
To properly perform the renormalization, it is standard to take the Minkowski spacetime limit by sending the plate separation to infinity, $L \to \infty$.  
Physically, this corresponds to removing the boundary conditions imposed by the plates and thus isolating the vacuum contributions intrinsic to free Minkowski space.  
This limit filters out finite-size effects, leaving only the divergent terms that require renormalization.  
Applying this limit to Eq.~\eqref{vj1}, we obtain
\begin{eqnarray}
V_{\text{eff}}\left( \Psi \right) &=&
\frac{m^{2}+C_{2}}{2}\,\Psi ^{2}
+ \frac{\lambda _{\psi} + C_{1}}{4!}\,\Psi ^{4}
+ C_{3}
+ \frac{M_{\lambda }^{4}}{64\pi ^{2}}
\left[ \ln\left( \frac{M^2_{\lambda }}{\alpha^2}\right) - \frac{3}{2}\right]
\notag \\
&& + \frac{(eB)^{2}}{4\pi^{2}}\,\zeta _{H}(-1,c) 
\left[ 1 - \ln\left( \frac{2eB}{\beta^{2}}\right)\right] 
+ \frac{(eB)^{2}}{4\pi^{2}} 
\left[ \frac{d}{ds}\zeta _{H}\left( s-1,c\right) \right]_{s=0}\,.
\label{effPLarge_L}
\end{eqnarray}

It is now convenient to express the Hurwitz zeta function and its derivative 
evaluated, respectively, at $s=-1$ and $s=0$ in terms of the parameter $c$ 
defined in Eq.~\eqref{c_coef}. From Ref.~\cite{elizalde1995zeta} we have
\begin{eqnarray}
\zeta _{H}\left( -1,c\right) 
&=&-\frac{1}{2}\left( c^{2}-c+\frac{1}{6}\right)  \notag \\
&=&\frac{1}{24}-\frac{M_{g}^{4}}{8\left( eB\right) ^{2}}.
\label{zetaH}
\end{eqnarray}
Furthermore, using the result from 
Refs.~\cite{elizalde2012ten,andrews1999special}, 
the derivative of the Hurwitz zeta function
evaluated at $s=0$ reads
\begin{eqnarray}
\left[ \frac{d}{ds}\zeta _{H}\left( s-1,c\right) \right]_{s=0}  
&=& \frac{1}{2} c(c - 1) \ln c - \frac{1}{4} c^2 + I(c) \nonumber\\
&=& -\zeta_{H}\left(-1,c\right)\ln c - \frac{1}{12}\ln c - \frac{1}{4} c^2 + I(c) ,
\label{derivative_zataH}
\end{eqnarray}
where 
\begin{eqnarray}
I(c) = \int_0^\infty 
\frac{ 2c\tan^{-1}(t/c) + t\ln\!\big(t^2 + c^2\big)}
     {e^{2\pi t}-1} \, dt,
\label{int}
\end{eqnarray}
which is finite for any value of $c$.

In order to eliminate the logarithmic dependence on $\beta$ 
present in Eq.~\eqref{effPLarge_L}, we take advantage of the fact that 
the natural scale of the system is set by $eB$ and choose 
$\beta^2 = 2 eB$. This removes the term proportional to 
$\ln(2 eB / \beta^2)$. 

Thus, combining Eqs.~\eqref{effPLarge_L}, \eqref{zetaH}, 
and \eqref{int}, we obtain
\begin{eqnarray}
V_{\text{eff}}\left( \Psi \right) &=&
\frac{\lambda _{\psi} + C_{1}}{4!}\,\Psi ^{4} 
+ \frac{m^{2}+C_{2}}{2}\,\Psi ^{2}
+ C_{3} \notag \\
&& + \frac{M_{\lambda }^{4}}{64\pi ^{2}}
\left[ \ln\left( \frac{M^2_{\lambda }}{\alpha^2}\right) - \frac{3}{2} \right]  
- \frac{3\mu^4\,c^2}{64\pi^2\left(c_0 - \frac{1}{2}\right)^2} \notag \\
&& - \frac{\mu^4}{32\pi^2\left(c_0 - \frac{1}{2}\right)^2}
\left[ \frac{1}{6} - c\big(1-\ln c\big) \right]\notag\\  
&&+ \frac{\mu^4}{16\pi^2\left(c_0 - \frac{1}{2}\right)^2}I(c) 
+ \frac{\mu^4\,c^2\ln c}{32\pi^2\left(c_0 - \frac{1}{2}\right)^2} ,
\label{Veff_L_large}
\end{eqnarray}
where $(eB)$ has been expressed in terms of $c_0$ written as follows,
\begin{eqnarray}
c_0 = \frac{1}{2} + \frac{\mu^2}{2 eB},
\label{c0}
\end{eqnarray}
which in fact corresponds to the definition of $c$ in Eq.~\eqref{c_coef} 
evaluated at $\Psi=0$. Note that the parameter $c_0$ encodes the effect of the external magnetic field $B$ 
on the spectrum of the system when the scalar field is in its ground state 
($\Psi = 0$). In the absence of the magnetic field, $c_0 \to \infty$, 
whereas for strong fields $eB \gg \mu^2$, $c_0$ approaches the minimal value 
$c_0 \to 1/2$. This behavior reflects the fact that $c_0$ is directly related 
to the Landau-level structure of the excitations and controls how the 
Hurwitz zeta-function contributions in Eqs.~\eqref{zetaH} and \eqref{derivative_zataH} 
affect the effective potential.

Let us now turn directly to the calculation of the renormalization constants $C_i$ $(i=1,2,3)$, which are essential to absorb the divergences and fix the physical parameters in our effective potential. We start by determining the renormalization coefficient $C_3$, associated with the vacuum energy normalization.  
This is done by imposing the physically motivated condition that the effective potential vanishes in the vacuum state, i.e., when the field expectation value is zero:
\begin{equation}
\lim_{\Psi \to 0} V_{\text{eff}}(\Psi) = 0.
\label{NC_3}
\end{equation}
Applying this condition to Eq.~\eqref{Veff_L_large} allows us to solve for $C_3$, ensuring the vacuum energy is properly normalized, i.e.,
\begin{align}
C_3 &=  \frac{m^{4}}{64\pi ^{2}}\left[ \frac{3}{2} - \ln\left( 
\frac{m^{2}}{\alpha ^{2}}\right) \right] + \frac{3 \mu^4 c_0^2}{64\pi^2 \left(c_0 - \frac{1}{2}\right)^2} \notag \\
&\quad + \frac{\mu^4}{32\pi^2 \left(c_0 - \frac{1}{2}\right)^2} \left[ \frac{1}{6} - c_0 (1 - \ln c_0) \right] \notag \\
&\quad - \frac{\mu^4}{16\pi^2 \left(c_0 - \frac{1}{2}\right)^2} I(c_0) - \frac{\mu^4 c_0^2 \ln c_0}{32\pi^2 \left(c_0 - \frac{1}{2}\right)^2}.
\label{coefficient_C_3}
\end{align}

Next, we obtain the coefficient $C_2$, which is responsible for the mass renormalization of the scalar field.  
To fix $C_2$, we require that the curvature of the effective potential at $\Psi = 0$ matches the physical squared mass $m^2$,
\begin{equation}
\left. \frac{d^{2} V_{\text{eff}}(\Psi)}{d\Psi^{2}} \right|_{\Psi=0} = m^{2}.
\label{RC2}
\end{equation}
Applying this condition to the non-renormalized effective potential \eqref{Veff_L_large} yields
\begin{align}
\frac{C_2}{2} &= \frac{m^{2} \lambda_{\psi}}{64 \pi^{2}} \left[ 1 + \ln \left( \frac{\alpha^{2}}{m^{2}} \right) \right] \notag \\
&\quad + \frac{g \mu^{2}}{32 \pi^{2} \left( c_0 - \frac{1}{2} \right)} \Big[ (1 - 2 c_0) \ln c_0 + 2 c_0 - 4 I_2(c_0) \Big],
\label{coefficient_C_2}
\end{align}
where
\begin{equation}
I_2(c_0) = \int_0^\infty dt \, \frac{\tan^{-1}(t / c_0)}{e^{2\pi t} - 1},
\label{int1}
\end{equation}
is a finite integral capturing nontrivial magnetic field contributions.

Finally, the coefficient $C_1$, which renormalizes the quartic coupling constant, is fixed by imposing that the fourth derivative of the effective potential at the origin equals the physical coupling $\lambda_{\psi}$,
\begin{equation}
\left. \frac{d^{4} V_{\text{eff}}(\Psi)}{d\Psi^{4}} \right|_{\Psi=0} = \lambda_{\psi}.
\label{RC1}
\end{equation}
This leads to
\begin{align}
\frac{C_1}{4!} &= \frac{\lambda_{\psi}^2}{256 \pi^{2}} \ln \left( \frac{\alpha^{2}}{m^{2}} \right) \notag \\
&\quad + \frac{g^{2}}{64 \pi^{2} c_0} \left[ 1 - 2 c_0 \ln c_0 + 4 c_0 I_1(c_0) \right],
\label{coefficient_C_1}
\end{align}
where
\begin{equation}
I_1(c_0) = \int_0^\infty dt \, \frac{t (t^{2} + c_0^{2})^{-1}}{e^{2\pi t} - 1},
\label{int2}
\end{equation}
is another finite integral encoding quantum corrections due to the external magnetic field.

\medskip

With the renormalization constants fixed as above, we combine all results to write the fully renormalized effective potential as
\begin{align}
V_{\rm eff}^{\text{R}}(\Psi) &= \frac{m^{2}}{2} \Psi^{2} + \frac{\lambda_{\psi}}{4!} \Psi^{4} + \frac{m^{4}}{64 \pi^{2}} \ln \left( \frac{M_{\lambda}^{2}}{m^{2}} \right) \notag \\
&\quad + \frac{\lambda_{\psi}^{2}}{256 \pi^{2}} \Psi^{4} \left[ \ln \left( \frac{M_{\lambda}^{2}}{m^{2}} \right) - \frac{3}{2} \right] + \frac{m^{2} \lambda_{\psi}}{64 \pi^{2}} \Psi^{2} \left[ \ln \left( \frac{M_{\lambda}^{2}}{m^{2}} \right) - \frac{1}{2} \right] \notag \\
&\quad + \frac{g^{2}}{64 \pi^{2} c_0} \Psi^{4} \left[ 1 - 2 c_0 \ln c_0 + 4 c_0 I_1(c_0) \right] \notag \\
&\quad + \frac{g \mu^{2}}{32 \pi^{2} \left( c_0 - \frac{1}{2} \right)} \Psi^{2} \left[ (1 - 2 c_0) \ln c_0 + 2 c_0 - 4 I_2(c_0) \right] \notag \\
&\quad + \frac{3 \mu^{4} (c_0^{2} - c^{2})}{64 \pi^{2} \left( c_0 - \frac{1}{2} \right)^{2}} + \frac{\mu^{4}}{32 \pi^{2} \left( c_0 - \frac{1}{2} \right)^{2}} \left[ (c - c_0) + (c_0 \ln c_0 - c \ln c) \right] \notag \\
&\quad + \frac{\mu^{4}}{16 \pi^{2} \left( c_0 - \frac{1}{2} \right)^{2}} \left[ I(c) - I(c_0) \right] + \frac{\mu^{4}}{32 \pi^{2} \left( c_0 - \frac{1}{2} \right)^{2}} \left[ c^{2} \ln c - c_0^{2} \ln c_0 \right] \notag \\
&\quad - \frac{M_{\lambda}^{2}}{8 \pi^{2} L^{2}} \sum_{n=1}^{\infty} n^{-2} K_2 \left( 2 n M_{\lambda} L \right) \notag \\
&\quad - \frac{1}{4 \pi^{2} L^{4}} \sum_{j=1}^{\infty} (-1)^j \mathcal{I}_0 (j, L, \Psi).
\label{ReffP}
\end{align}
This expression represents the renormalized effective potential, which incorporates vacuum energy normalization, mass and coupling renormalization, and finite-size and topological corrections due to the plates. It forms the basis for subsequent calculations of the vacuum energy per unit area of the plates, its first-order coupling-constant corrections, and the topological mass in the system.

It is worth emphasizing that, in the present analysis, we have adopted
renormalization conditions that do not require the magnetic field to vanish.
Instead, our procedure removes the divergent terms by subtracting the
corresponding Minkowski spacetime contribution (no plates), without imposing
additional constraints on the finite part of the magnetic field. This choice
is consistent with a physical scenario in which the field can extend beyond
the region between the plates, allowing for non-vanishing values outside the
bounded domain.

Alternatively, one could impose the renormalization condition that the magnetic
field vanishes identically. This would correspond to a situation in which the
field is strictly confined to the inter-plate region, leading to different
finite contributions to the physical quantities of interest \cite{bordag2009advances, milton2001casimir}.
Renormalization conditions
requiring the magnetic field \(B\) to vanish in the vacuum state results in a renormalized effective potential whose vacuum energy,
evaluated at \(\Psi=0\), still contained residual terms proportional to \(B^{2}\)
that did not depend on the plate separation \(L\).

These \(L\)-independent terms imply a vacuum energy offset that persists even in
the limit \(L \to \infty\), contradicting the physical expectation that the vacuum
energy should vanish when the plates are infinitely separated and the system
recovers free Minkowski spacetime. Such residual contributions can be interpreted
as artifacts of the renormalization scheme rather than physical effects. By contrast, the renormalization conditions adopted in the present work do not
impose the vanishing of \(B\) in the vacuum. Instead, divergences are subtracted
using the Minkowski spacetime limit without plates, allowing the magnetic field
to remain finite. This choice ensures that the vacuum energy naturally tends to zero
as \(L \to \infty\), providing a physically consistent baseline and avoiding
spurious constant or linear terms in \(L\). Thus, the presence of residual \(B^{2}\)-dependent vacuum energy terms in
the renormalization scheme that forces \(B = 0\) highlights the importance of
carefully selecting renormalization conditions that respect the physical context
and boundary conditions of the system.

Finally, it is important to note that there are physical situations in which imposing 
renormalization conditions at vanishing magnetic field, \(B=0\), is both 
natural and physically justified. A typical example is provided by the 
calculation of magnetic-field effects on atomic systems in the presence of a 
uniform, external, and non-confined field. In such a scenario, the reference 
vacuum is naturally taken to be the free-space vacuum with \(B=0\), and the 
magnetic field is treated as an external perturbation that permeates all space, 
without altering the topology or the mode structure of the system. This 
approach was followed, for instance, in the work of Erdas and Feldman 
\cite{Erdas:1990gy}.

\section{Vacuum Energy, Topological Mass and First-Order Coupling-Constant Correnctions}
\label{secIV}
\subsection{Vacuum energy per unit area of the plates}
\label{subIVA}
Before calculating the vacuum energy per unit area of the plates, let us first analyze the renormalized effective potential given in Eq.~\eqref{ReffP} evaluated at $\Psi=0$, that is,
\begin{eqnarray}
V_{\rm eff}^{\text{R}}(0) &=& -\frac{m^{2}}{8\pi^{2}L^{2}} \sum_{n=1}^{\infty} n^{-2} K_{2}\left( 2 n m L \right) \notag \\
&& - \frac{1}{4\pi^{2} L^{4}} \sum_{j=1}^{\infty} (-1)^j \mathcal{I}_0(j, L, 0) \,,
\label{ReffPvacuum}
\end{eqnarray}
where the integral function $\mathcal{I}_0(j, L, 0)$ is defined as
\begin{equation}
\mathcal{I}_0(j, L, 0) = B_L^2 \int_0^{\infty} d\tau \, \frac{\tau^{-3}}{\sinh(B_L^2 \tau^2)} \, e^{-\mu_L^2 \tau^2 - \frac{j^2}{\tau^2}},
\label{integral_B0_zero}
\end{equation}
with $\mu_L = \mu L$ and $B_L$ related to the magnetic field parameters as defined previously in Eq.~\eqref{c_coef}.

The vacuum energy per unit area of the plates can then be defined as
\begin{eqnarray}
\frac{E_{\rm R}}{A} = L V_{\rm eff}^{\text{R}}(0).
\label{E_per_area}
\end{eqnarray}
It is worth emphasizing that the effective potential $V_{\rm eff}$ is naturally defined as the vacuum energy density, i.e., the energy per unit volume of the system. Since the geometry considered consists of two parallel plates separated by a distance $L$, the physical volume relevant to the problem is the area of the plates multiplied by this separation. Therefore, to obtain the vacuum energy per unit area of the plates, it is necessary to multiply the effective potential by $L$. This operation converts the volume energy density into an energy per unit area, allowing a direct comparison with the results obtained via the mode summation method, which usually provides the vacuum energy per unit area from the outset \cite{bordag2009advances, milton2001casimir}. In other words, the multiplication by $L$ ensures consistency between the two formalisms and reflects the geometric configuration of the system under consideration.
\begin{figure}[htbp]
    \centering
    \includegraphics[width=0.53\textwidth]{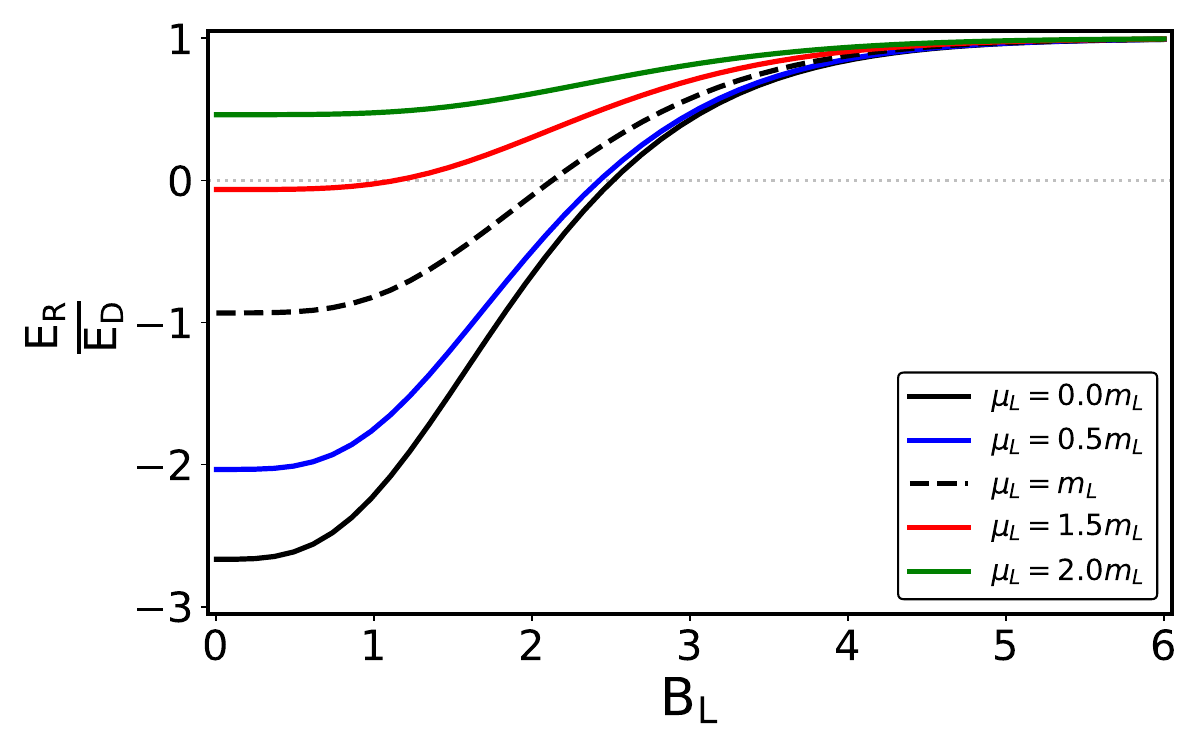} 
    \caption{Ratio $\frac{E_R}{E_D}$ obtained from Eq.~\eqref{VErenormalized} plotted against $B_L$ for fixed $mL=1$, denoted as $m_L$. The curves illustrate the influence of the magnetic field contribution on the total vacuum energy.}
    \label{fig2}
\end{figure}

From Eqs.~\eqref{ReffPvacuum} and \eqref{E_per_area} we, thus, have
\begin{eqnarray}
\frac{E_{\rm R}}{A} = &-& \frac{m^{2}}{8 \pi^{2} L} \sum_{n=1}^{\infty} n^{-2} K_{2}(2 n m L)\notag\\
&-& \frac{1}{4 \pi^{2} L^{3}} \sum_{j=1}^{\infty} (-1)^j \mathcal{I}_0(j, L, 0) \,.
\label{VErenormalized}
\end{eqnarray}
Here, the first term on the r.h.s. corresponds to the vacuum energy contribution of a real scalar field obeying Dirichlet boundary conditions, which we will denote by $\frac{E_D}{A}$. The second term represents the contribution from a charged complex scalar field subject to mixed boundary conditions in the presence of a magnetic field, denoted by $\frac{E_B}{A}$.

\begin{figure}[htbp]
    \centering
    \includegraphics[width=0.495\textwidth]{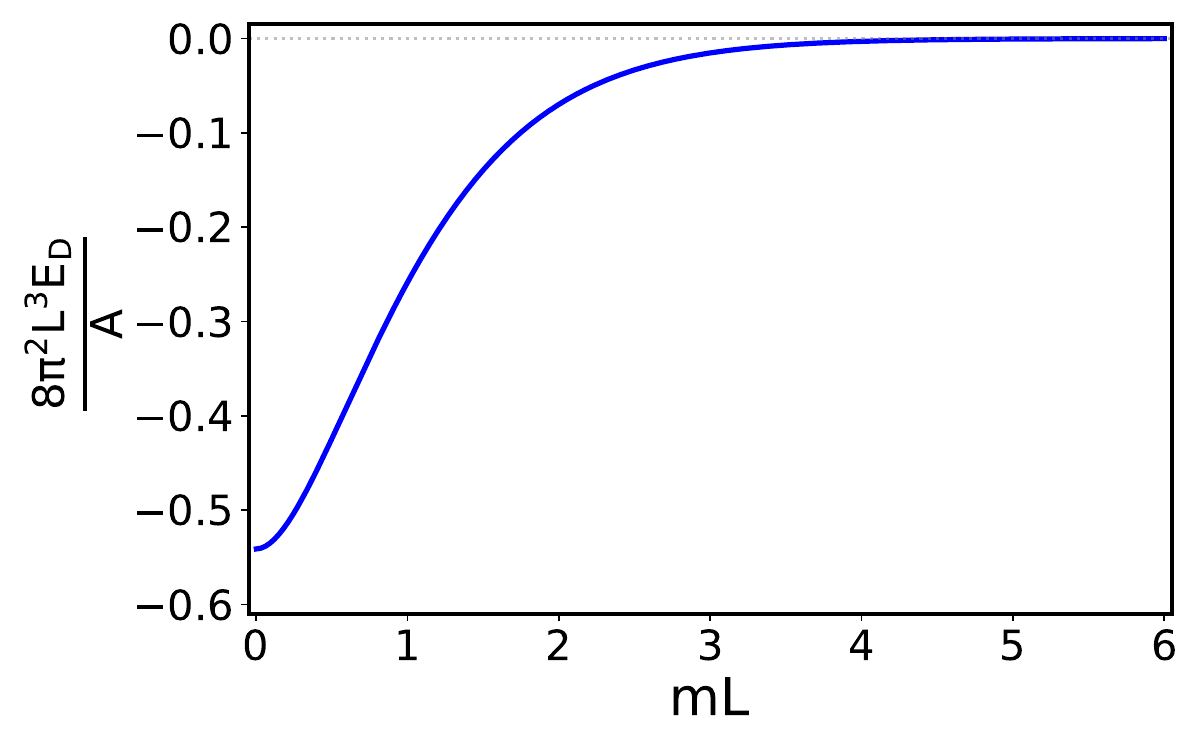} 
    \hspace{-0.2cm}
    \includegraphics[width=0.5\textwidth]{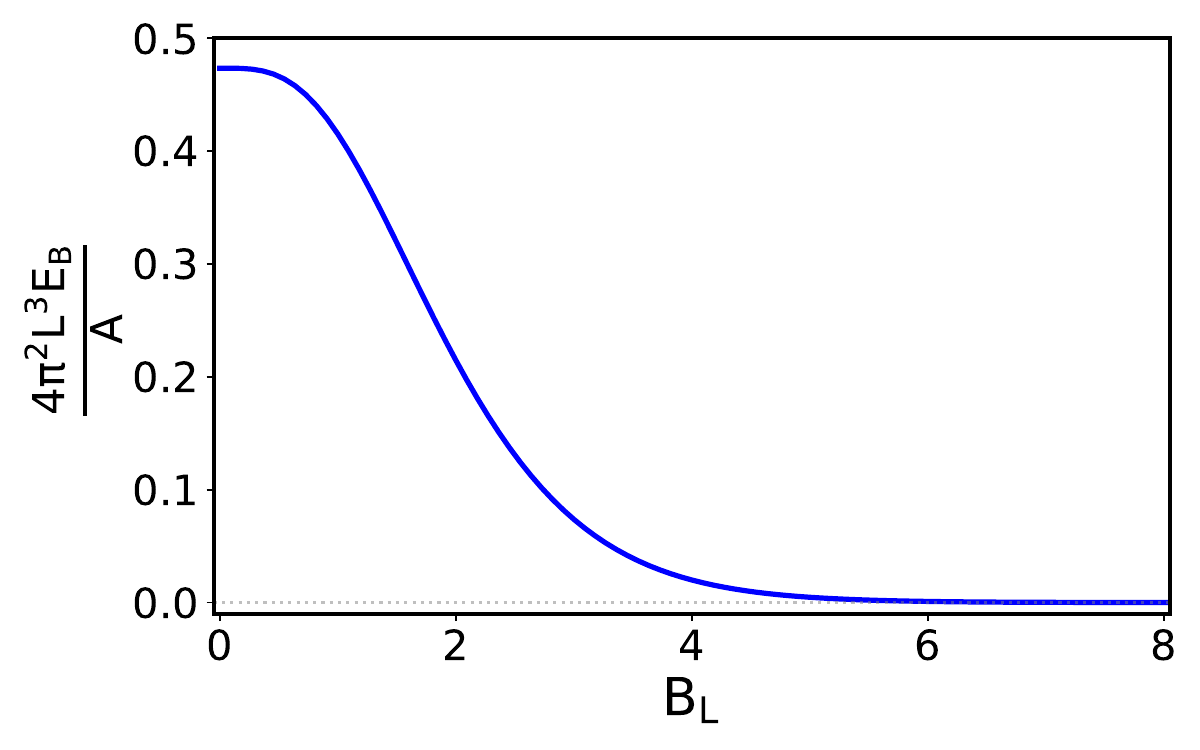} 
    \caption{Vacuum energy contributions per unit area for Dirichlet part $\frac{E_D}{A}$ (left) and massless magnetic part $\frac{E_B}{A}$ (right). These plots illustrate the individual roles of boundary conditions and magnetic effects on the vacuum energy.}
    \label{fig3}
\end{figure}

Fig.~\ref{fig2} shows the ratio $\frac{E_R}{E_D}$ as a function of the dimensionless parameter $B_L$, for a fixed mass parameter $mL=1$. This ratio highlights the role of the magnetic field contribution in modifying the vacuum energy relative to the purely Dirichlet case. In particular, the presence of the magnetic field can determine whether the total vacuum energy becomes positive or remains negative. It is worth noting that the Dirichlet contribution $\frac{E_D}{A}$ is itself strictly negative, as illustrated by the plot on the left panel of Fig.~\ref{fig3}.

\begin{figure}[htbp]
    \centering
    \includegraphics[width=0.6\textwidth]{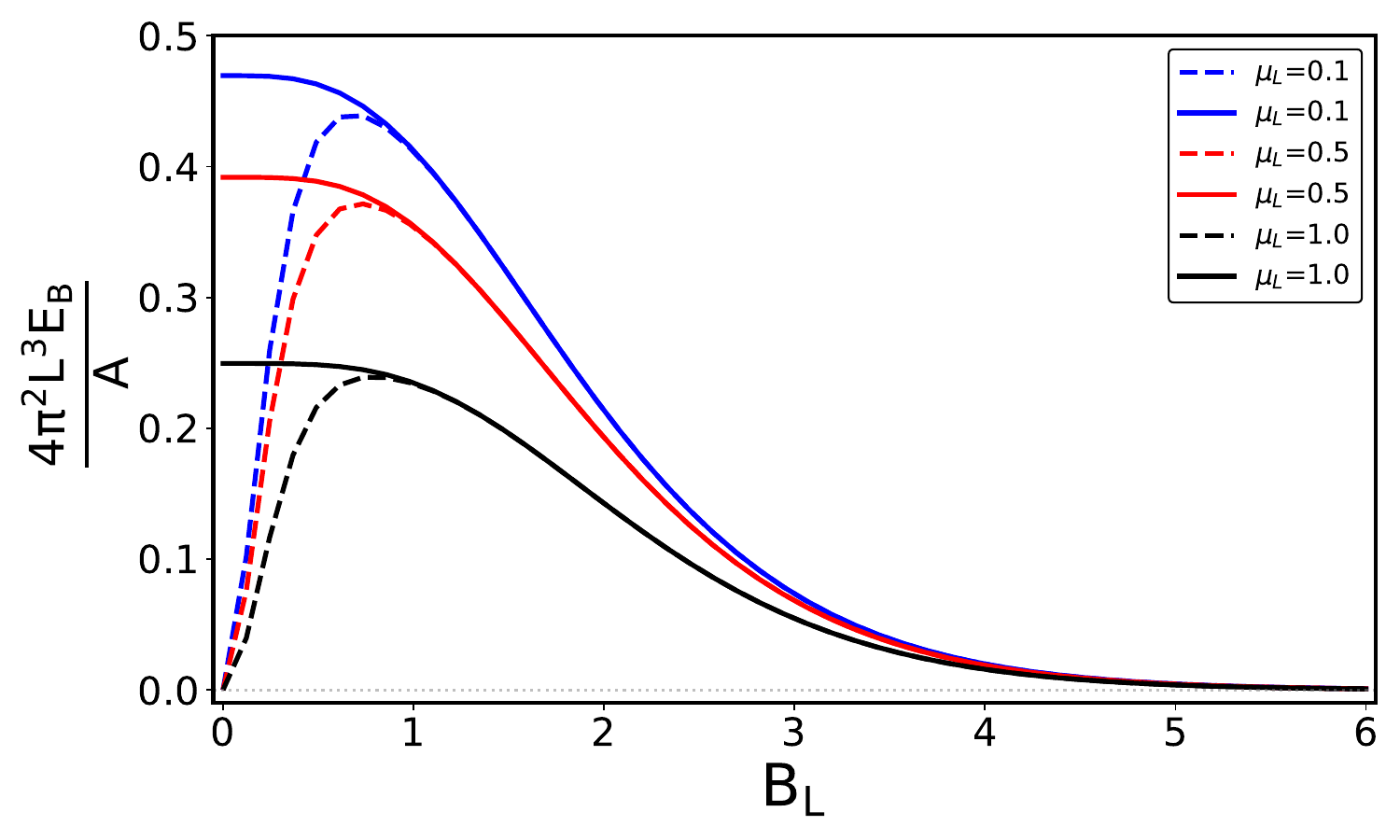} 
    \caption{Comparison between the expressions for $\frac{E_B}{A}$ given in Eqs.~\eqref{VErenormalized_alternative} and \eqref{VErenormalized}, for $\mu_L=0.1$, 0.5 and 1.0.}
    \label{fig4}
\end{figure}

For further insight, Fig.~\ref{fig3} displays separately the vacuum energy contributions $\frac{E_D}{A}$ (left panel) and $\frac{E_B}{A}$ (right panel), being this latter for the massless case. The behavior of these contributions as functions of the plate separation, $L$, reveals their distinct scaling properties and physical origins. The corresponding plot for the massive case of $\frac{E_B}{A}$ is shown later in Fig.~\ref{fig4}.

In summary, the total vacuum energy density in our system arises from two distinct physical sources; the quantized real scalar field confined by Dirichlet boundaries, and the charged complex scalar field influenced by the external magnetic field and mixed boundary conditions. Understanding how these contributions combine is crucial for further analysis of the first-order coupling-constant corrections and the emergence of a topological mass in the theory.

To complete our analysis, we now consider the asymptotic limits of the vacuum energy in the regimes where the magnetic field is either weak or strong. The asymptotic behavior of the Dirichlet contribution has already been investigated in several works in the literature (see, for instance, Refs.~\cite{bordag2009advances} and \cite{Cruz:2020zkc}). Here, our focus is to provide a more detailed study of the asymptotic limits associated with the constant magnetic field contribution.

We start with the strong-field limit, $B_L \gg 1$, of the vacuum energy $\frac{E_B}{A}$. At leading order, it can be obtained by using the following approximation for Eq.~\eqref{integral_B0_zero}:
\begin{eqnarray}
\mathcal{I}_0(j, L, 0) &\simeq& 2 B_L^2 \int_{0}^{\infty} d\tau \, \tau^{-3} \, e^{-\left(\mu_L^2 + B_L^2\right)\tau^2 - \frac{j^2}{\tau^2}} \notag \\
&\simeq& z B_L^2 \, \frac{K_1(j z)}{j},
\label{integral_B0approx}
\end{eqnarray}
where we have defined $z = 2\sqrt{B_L^2 + \mu_L^2}$, and in this limit we have also used the approximation $\sinh(B_L^2 \tau^2) \simeq \frac{1}{2} e^{B_L^2 \tau^2}$. Consequently, by making use of the asymptotic limit for large arguments of the Macdonald function, i.e., $K_{\nu}(z)\simeq\sqrt{\frac{\pi}{2z}}e^{-z}$, we obtain \cite{abramowitz1965handbook}
\begin{eqnarray}
\frac{E_{\rm B}}{A} &\simeq& \frac{B_L^2 z^{\frac{1}{2}}}{2^{\frac{5}{2}} \pi^{\frac{3}{2}} L^3} \, e^{-z} \notag \\
&\simeq& \frac{B_L^{\frac{5}{2}}}{4 \pi^{\frac{3}{2}} L^3} \, e^{-2 B_L} \,,
\label{VErenormalized_strong}
\end{eqnarray}
where the leading contribution corresponds to $j = 1$, and we have also imposed the additional requirement $B_L \gg \mu_L$.

On the other hand, in the weak-field limit, $B_L \ll 1$, Eq.~\eqref{integral_B0_zero} can be expressed in the form
\begin{eqnarray}
\mathcal{I}_0(j, L, 0)  = 2B_L^2\int_{0}^{\infty} \frac{d\tau\,\tau^{-3}}{1 - e^{-2B_L^2\tau^2}} 
\, e^{-(\mu_L^2 + B_L^2)\tau^2 - \frac{j^2}{\tau^2}}.
\label{integral_B0_zero10}
\end{eqnarray}
Next, we consider the Taylor expansion
\begin{eqnarray}
\frac{1}{{1 - e^{-2w}} } \simeq \frac{1}{2w} + \frac{1}{2} + \frac{w}{6} - \frac{w^3}{90} + \mathcal{O}(w^5).
\label{series}
\end{eqnarray}
From Eqs.~\eqref{integral_B0_zero10} and \eqref{series}, the complex-field contribution to the vacuum energy in Eq.~\eqref{VErenormalized}, within the weak-field approximation and in dimensionless form, can be written as
\begin{eqnarray}
\frac{4\pi^2L^3E_{\rm B}}{A}&\simeq&- z^2\sum_{j=1}^{\infty}(-1)^j\frac{K_2(zj)}{4j^2} -
 zB_L^2\sum_{j=1}^{\infty}(-1)^j\frac{K_1(zj)}{2j}\notag\\
 &&- \frac{B_L^4}{3}\sum_{j=1}^{\infty}(-1)^jK_0(zj) 
 + \frac{4B_L^8}{45z^2}\sum_{j=1}^{\infty}(-1)^j j^2K_2(zj).
\label{VErenormalized_weak}
\end{eqnarray}
Taking $eB\to 0$ directly in the first term recovers the known result for pure mixed boundary condition. This provides an alternative form to the expression obtained in Ref.~\cite{Cruz:2020zkc}, where a real scalar field was considered. 

We can further consider the limit $z\ll 1$ in Eq.~\eqref{VErenormalized_weak} by expanding the Macdonald functions. Before doing so, we first perform the sum over $j$. For this purpose, we employ the following integral representation of the Macdonald function~\cite{abramowitz1965handbook}:
\begin{eqnarray}
K_{\nu}(jz)=\int_{0}^{\infty}e^{-jz\cosh(t)}\cosh(\nu t)\,dt.
\label{representation2}
\end{eqnarray}

For the second term in Eq.~\eqref{VErenormalized_weak}, using Eq.~\eqref{representation2} and performing the sum over $j$, we obtain 
\begin{eqnarray}
T_2&=&\frac{B_L^2z}{2}\int_{0}^{\infty}\ln\!\big(1 + e^{-z\cosh(t)}\big)\cosh(t)\,dt\notag\\
&=&\frac{B_L^2z}{2}\int_{1}^{\infty}\frac{u\,\ln\!\big(1 + e^{-zu}\big)}{(u^2 - 1)^{\frac{1}{2}}}du\notag\\
&\simeq&\frac{B_L^2z}{2}\int_{1}^{\infty}\ln\!\big(1 + e^{-zu}\big)\,du,
\label{term2}
\end{eqnarray} 
where the change of variables $u=\cosh(t)$ has been applied. In the regime $z\ll 1$, the upper limit, $u\gg 1$, dominates the integral. Performing the integration and expanding for small $z$, we find
\begin{eqnarray}
T_2&\simeq&\frac{B_L^2z}{2}\left(\frac{\pi^2}{12z} - \ln 2 + \frac{z}{4} + O(z^2)\right).
\label{term21}
\end{eqnarray} 

For the third term in Eq.~\eqref{VErenormalized_weak}, setting $\nu=0$ in the Macdonald function of Eq.~\eqref{representation2} and summing over $j$ yields
\begin{eqnarray}
T_3=\frac{B_L^4}{3}\int_{0}^{\infty}\frac{e^{-z\cosh(t)}}{1+ e^{-z\cosh(t)}}\,dt.
\label{term3}
\end{eqnarray} 
When $z\ll 1$, the integral is dominated by the lower limit, $t\ll 1$. We can then use the approximation,
\begin{eqnarray}
T_3&\simeq&\frac{B_L^4}{3}\frac{1}{1+e^{-z}}\int_{0}^{\infty}e^{-z\cosh(t)}\,dt\notag\\
&\simeq&\frac{B_L^4}{3}\frac{K_0(z)}{1+e^{-z}}\notag\\
&\simeq&\frac{B_L^4}{3}\left(-\frac{\ln z}{2} + O(1)\right).
\label{term31}
\end{eqnarray}

Similarly, for the fourth term in Eq.~\eqref{VErenormalized_weak}, using Eq.~\eqref{representation2} and summing over $j$ gives 
\begin{eqnarray}
T_4=-\frac{4B_L^8}{45z^2}\int_{1}^{\infty}\frac{\left(e^{-z\cosh(t)} - e^{-2z\cosh(t)}\right)\cosh(2t)}{\left(e^{-z\cosh(t)} + 1\right)^3}dt.
\label{term4}
\end{eqnarray} 
For $z\ll 1$, this is dominated by the lower limit, so we have 
\begin{eqnarray}
T_4&\simeq&-\frac{4B_L^8}{45z^2}\frac{\left[K_2(z) - K_2(2z)\right]}{\left(e^{-z} + 1\right)^3}\notag\\
&\simeq&-\frac{4B_L^8}{45z^2}\left(\frac{3}{16z^2} + O(1/z)\right).
\label{term41}
\end{eqnarray} 

Substituting Eqs.~\eqref{term21}, \eqref{term31} and \eqref{term41} into Eq.~\eqref{VErenormalized_weak}, we obtain
\begin{eqnarray}
\frac{4\pi^2L^3E_{\rm B}}{A}&\simeq&- \mu_L^2\sum_{j=1}^{\infty}\frac{(-1)^j}{j^2}K_2(2j\mu_L) +\frac{\pi^2B_L^2}{24} - \frac{zB_L^2\ln 2}{2} \notag\\
 &&+ \frac{z^2B_L^2}{8}  -\frac{B_L^4}{6}\ln z - \frac{3B_L^8}{180z^4}.
\label{VErenormalized_weak2}
\end{eqnarray}

For the massless case $\mu_L\to 0$, we have
\begin{eqnarray}
\frac{4\pi^2L^3E_{\rm B}}{A}&\simeq& \frac{7\pi^4}{1440} + 
 \frac{\pi^2B_L^2}{24} - B_L^3\ln 2 - \frac{B_L^4}{6}\ln(2B_L) + \frac{479B_L^4}{960},
 \label{VErenormalized_weak_massless}
\end{eqnarray}
where, for the first term in Eq.~\eqref{VErenormalized_weak2}, we used the limit~\cite{abramowitz1965handbook}
\begin{eqnarray}
\lim_{x\to 0}x^{\nu}K_{\nu}(bx)= \left(\frac{2}{b}\right)^{\nu}\frac{\Gamma(\nu)}{2}.
\label{limit}
\end{eqnarray}
The first term in Eq.~\eqref{VErenormalized_weak_massless} is exactly twice the result obtained in Ref.~\cite{Cruz:2020zkc} for a real scalar field, as expected. 

For completeness, Eq.~\eqref{integral_B0_zero} can also be written in an alternative form by employing Eq.~\eqref{csch_expansion}, yielding
\begin{eqnarray}
\mathcal{I}_0(j, L, 0)  &=& 2B_L^2\sum_{n=0}^{\infty}\int_{0}^{\infty}d\tau\,\tau^{-3}e^{-\big(\mu_L^2 + (2n+ 1)B_L^2\big)\tau^2 - \frac{j^2}{\tau^2}}\notag\\
&=&2B_L^2\sum_{n=0}^{\infty}\mu_n\frac{K_1\big(2j\mu_n\big)}{j},
\label{integral_B0_zero1}
\end{eqnarray}
where $\mu_n = \sqrt{(2n + 1)B_L^2 + \mu_L^2}$. Consequently, the complex-field contribution reads 
\begin{eqnarray}
\frac{E_{\rm B}}{A} = -\frac{B_L^2}{2\pi^2L^3}\sum_{j=1}^{\infty}(-1)^j\sum_{n=0}^{\infty}\mu_n\frac{K_1\big(2j\mu_n\big)}{j}.
 \label{VErenormalized_alternative}
\end{eqnarray}
This expression has been previously obtained in Ref.~\cite{Sitenko:2014wwa} in a different context. Also, in other contexts, similar expressions to the one above have been derived in Refs.~\cite{Elizalde:2002kb, Cougo-Pinto:1998jun}, where the authors impose Dirichlet boundary conditions on the gauge-coupled complex scalar field. 

Note that both expressions in Eqs.~\eqref{VErenormalized} and \eqref{VErenormalized_alternative} represent the same physical quantity, i.e., the vacuum energy contribution from a charged scalar field under mixed boundary conditions in the presence of a uniform magnetic field. However, their practical validity over the entire range of magnetic field strengths differs due to the distinct ways they organize and encode the spectral information.

In Fig.~\ref{fig4}, we have comparatively plotted the expressions for $\frac{E_{\rm B}}{A}$ in Eqs.~\eqref{VErenormalized} (dashed lines) and \eqref{VErenormalized_alternative} (solid lines), for different values of $\mu_L$. As illustrated in the figure, the expression in Eq.~\eqref{VErenormalized_alternative} is accurate approximately for $B_L \gtrsim 1$, beyond which it converges rapidly and correctly accounts for the Landau-level quantization. Below this regime, the series expansion used in Eq.~\eqref{VErenormalized_alternative} becomes ill-defined. 

Physically, this means that Eq.~\eqref{VErenormalized_alternative} inherently assumes a sufficiently strong magnetic field to resolve discrete Landau levels, while Eq.~\eqref{VErenormalized} remains valid for all magnetic field strengths, including the weak-field regime where the spectrum is effectively continuous. Thus, the different domains of validity reflect the distinct spectral decompositions encoded by these expressions.

It should also be pointed out that the difference in the range of validity between the two expressions in Eqs.~\eqref{VErenormalized} and \eqref{VErenormalized_alternative}, for the vacuum energy contribution $\frac{E_B}{A}$, is fundamentally connected to the convergence properties of the series expansion in Eq.~\eqref{csch_expansion}, where in our case the argument is given by $T = B_L^2 \tau^2$. This expansion is valid and convergent only for strictly positive values of $T$. Since $B_L$ is proportional to the magnetic field strength, this implies that the series representation assumes a nonzero magnetic field. Physically, this corresponds to the well-known Landau level quantization in the presence of a magnetic field, where the discrete energy levels arise naturally.

However, when the magnetic field becomes very weak ($B_L \to 0$), the parameter $T$ approaches zero for
all values of the integration variable $\tau$. In this limit, the discrete Landau levels become densely spaced and effectively merge into a continuous spectrum. Consequently, the series expansion converges very slowly, making the sum representation ill-defined or numerically unstable in this regime.

On the other hand, the integral representation of $\mathcal{I}_0(j, L, 0)$ in Eq.~\eqref{VErenormalized} for the $\frac{E_B}{A}$ part, retains the hyperbolic sine function in the denominator without expanding it into a series. This avoids any implicit restrictions on $B_L$, allowing the integral form to be valid and well-defined for all values of the magnetic field strength, including the weak-field limit $B_L \to 0$.

In summary, the key physical reason why the sum over Landau levels expression is not valid for the entire range of $B_L$ lies in the convergence domain of the series expansion for $1/(2\sinh(T))$. The integral representation circumvents this limitation, smoothly describing the transition from discrete Landau levels at strong magnetic fields to a continuous spectrum at weak fields.

\subsection{Topological mass}
\label{secIVB}
The topological mass can be defined through the second derivative, with respect to $\Psi$, of the renormalized effective potential taken at the vacuum state, $\Psi=0$, namely
\begin{eqnarray}
m^2_{\rm T}=\frac{d^2V_{\rm eff}^{\text{R}}(\Psi)}{d\Psi^2}\Big|_{\Psi=0}.
\label{TMC}
\end{eqnarray}
By using the above definition in Eq.~\eqref{ReffP}, the topological mass naturally separates into three contributions, that is,
\begin{eqnarray}
m^2_{\rm T} = m^2 + m_{\rm D}^2 + m_{\rm B}^2,
\label{TM}
\end{eqnarray}
where \(m^2\) is the bare mass parameter, \(m_{\rm D}^2\) is the purely Dirichlet contribution, and \(m_{\rm B}^2\) encodes the correction induced by the external magnetic field and mixed boundary conditions.

In explicit form, one finds
\begin{eqnarray}
m^2_{\rm T} &=& m^2 + \frac{\lambda_{\psi}m}{8\pi^2L}\sum_{n=1}^{\infty}n^{-1}K_1\big(2mLn\big)\notag \\ 
&&+\frac{g}{2\pi^2L^2}\sum_{j=1}^{\infty}(-1)^{j}\,\mathcal{I}_{1}(j, L,0),
\label{TM}
\end{eqnarray}
where
\begin{eqnarray}
\mathcal{I}_{1}(j, L,0) = B_L^2\int_{0}^{\infty} \frac{d\tau\,\tau^{-1}}{\sinh(B_L^2\tau^2)} \,
e^{-\mu_L^2\tau^2 - \frac{j^2}{\tau^2}}.
\label{integral_B_topological}
\end{eqnarray}
\begin{figure}[htbp]
    \centering
    \includegraphics[width=0.5\textwidth]{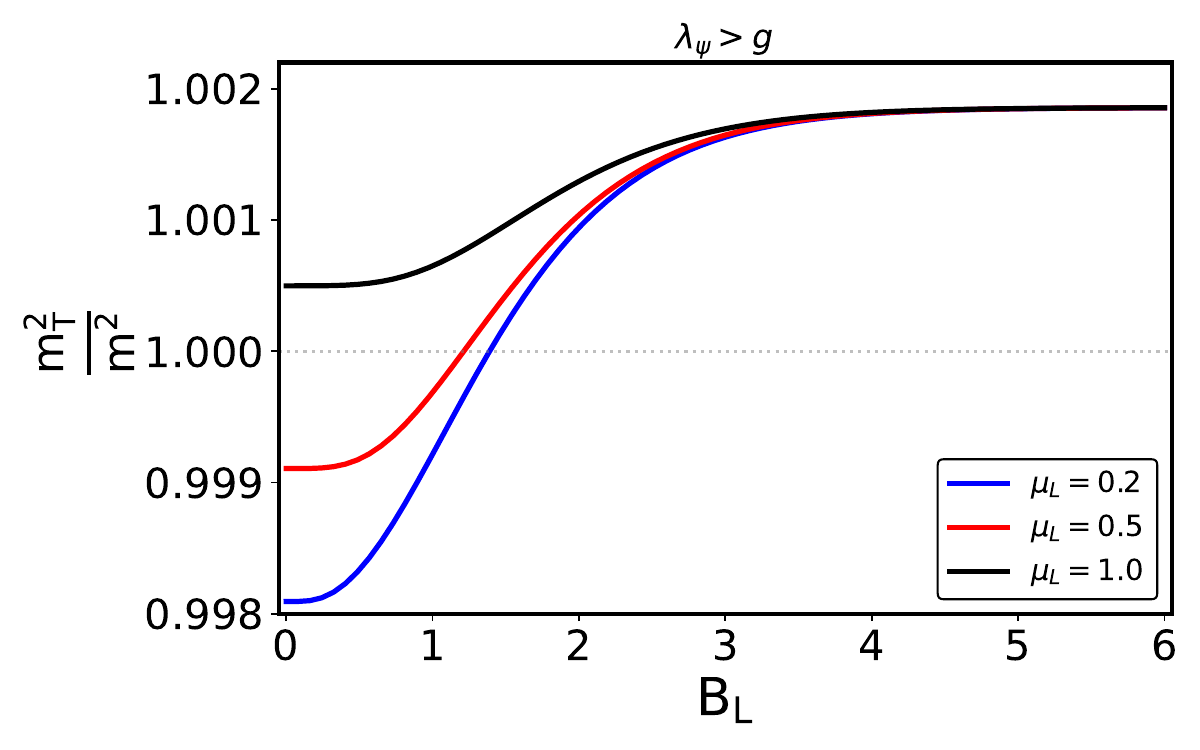} 
    \hspace{-0.2cm}
    \includegraphics[width=0.5\textwidth]{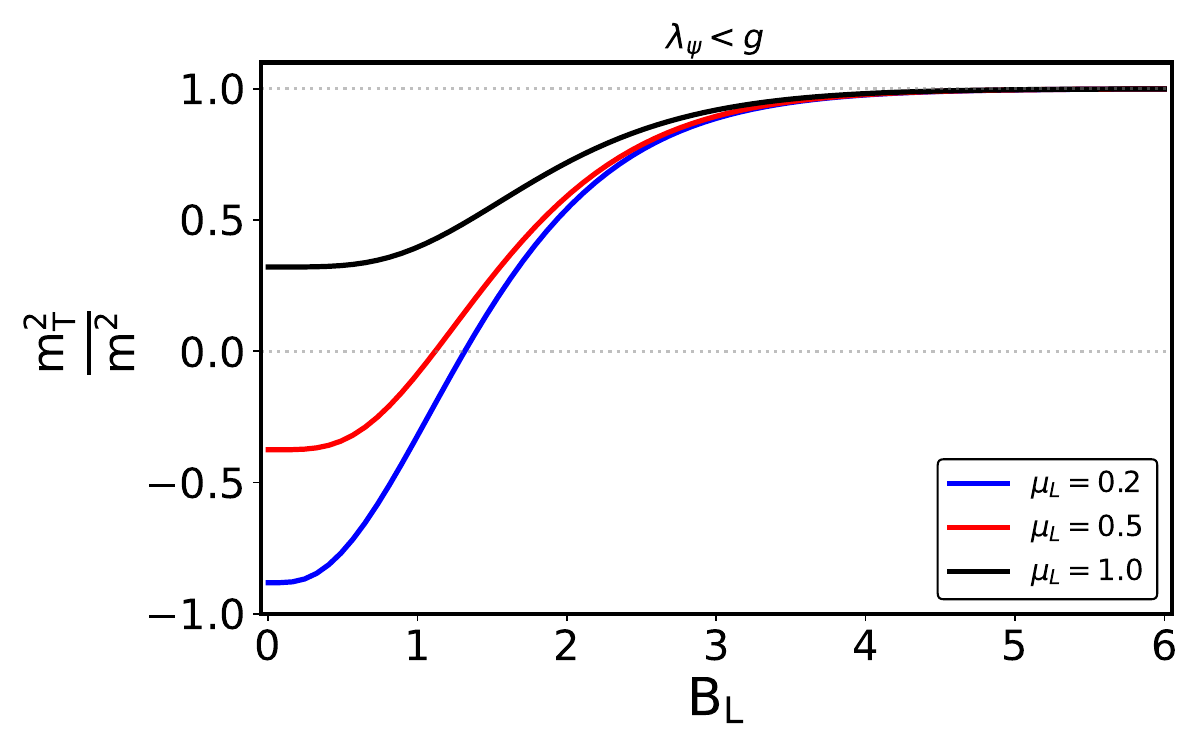}
    \caption{Topological mass given by Eq.~\eqref{TM} as a function of $B_L$, for $mL = 1$ and different values of $\mu_L$. The plots correspond to the cases $\lambda_{\psi} = 1.0$, $g = 0.2$ (left) and $\lambda_{\psi} = 0.02$, $g = 100$ (right).
}
    \label{fig5}
\end{figure}

The structure of Eq.~\eqref{TM} shows that \(m_{\rm T}^2\) depends explicitly on the coupling constants \(\lambda_{\psi}\) and \(g\), whose relative magnitudes can significantly influence the sign of the total mass squared. This opens the possibility of vacuum instabilities or symmetry breaking phenomena depending on the physical regime considered. A systematic vacuum stability analysis could be performed to fully explore the allowed parameter space, as it has been done in Refs. \cite{toms1980interacting, PhysRevD.107.125019}. However, in the present work we restrict ourselves to illustrating the qualitative behavior in relevant limits, leaving a detailed stability study for future work.

In Fig.~\ref{fig5}, we plot, in dimensionless form, the topological mass given by Eq.~\eqref{TM} as a function of $B_L$ for different values of $\mu_L$, with $mL = 1$. The plot on the right, corresponding to $\lambda_{\psi} < g$, shows that the topological mass can become negative. In contrast, for $\lambda_{\psi} > g$ (left), the topological mass remains positive.

When \(m=0\), Eq.~\eqref{TM} reduces, with the help of Eq.~\eqref{limit}, to the massless expression for the topological mass, given by
\begin{eqnarray}
m^2_{\rm T} &=&\frac{\lambda_{\psi}}{96L^2} + \frac{g}{2\pi^2L^2}\sum_{j=1}^{\infty}(-1)^{j}\,\mathcal{I}_{1}(j, L,0).
\label{TM_massless}
\end{eqnarray}
From the above expression, it is straightforward to see that, in order for the topological mass to be positive, one must have \(\lambda_{\psi} > g\), since the second term is the one that can become negative.
\begin{figure}[htbp]
    \centering
    \includegraphics[width=0.5\textwidth]{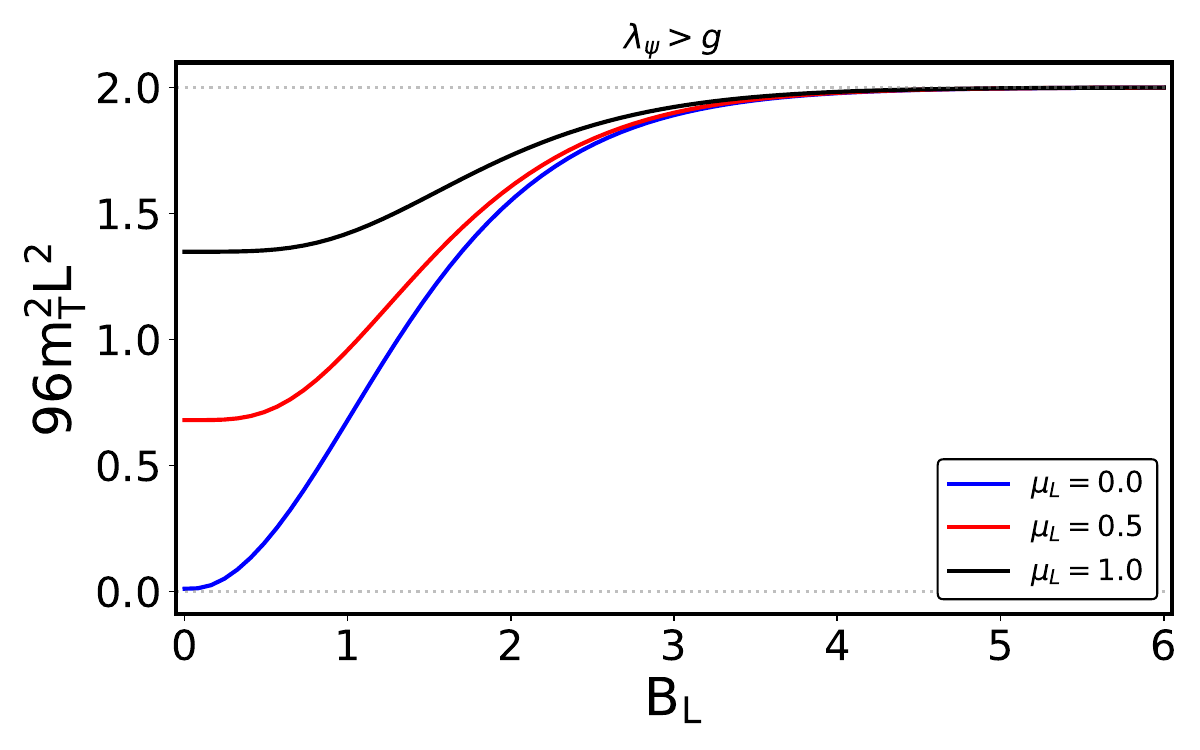} 
    \hspace{-0.2cm}
    \includegraphics[width=0.5\textwidth]{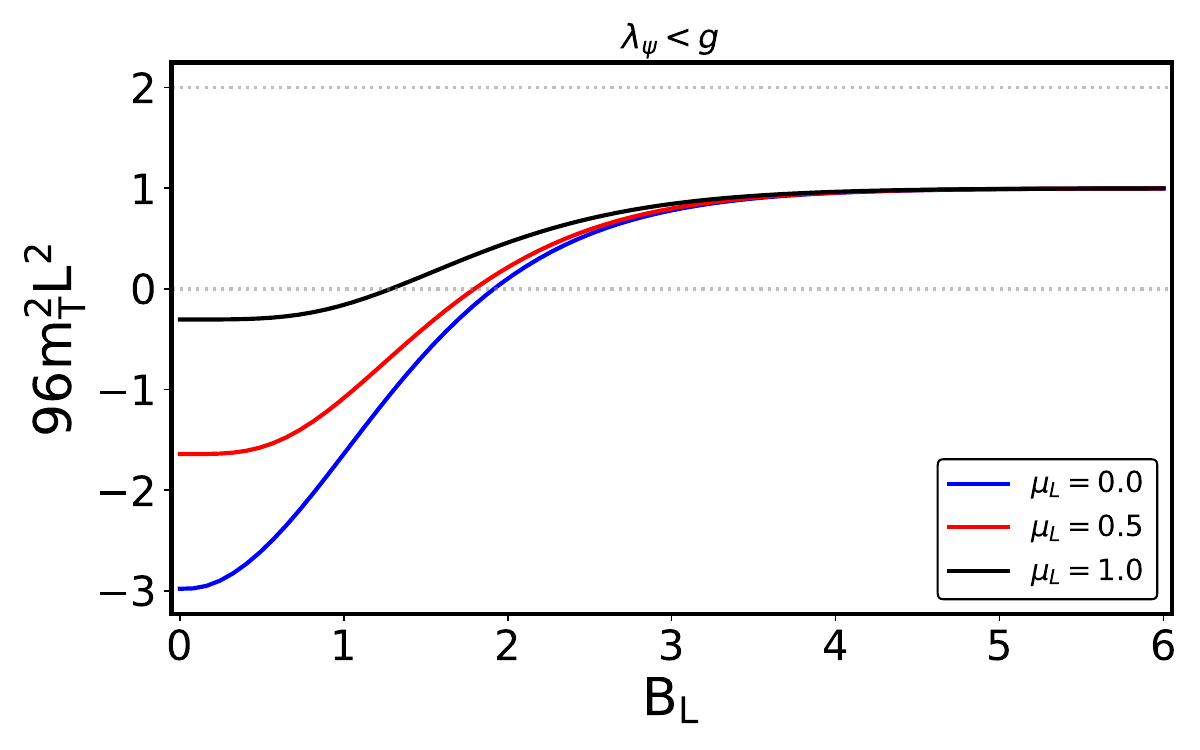}
    \caption{Topological mass given by Eq.~\eqref{TM_massless} as a function of $B_L$, and different values of $\mu_L$. The plots correspond to the cases $\lambda_{\psi} = 2.0$, $g = 1.0$ (left) and $\lambda_{\psi} = 1.0$, $g = 2.0$ (right).
}
    \label{fig6}
\end{figure}

In Fig.~\ref{fig6}, we plot, in dimensionless form, the topological mass given by Eq.~\eqref{TM_massless} as a function of $B_L$ for different values of $\mu_L$. The plot on the right, corresponding to $\lambda_{\psi} < g$, shows that the massless topological mass can become negative. In contrast, for $\lambda_{\psi} > g$ (left), the topological mass remains positive.

Once again, we want to focus on the analysis of the asymptotic limits when the magnetic field can be either weak or strong. The analysis for the case of Dirichlet boundary conditions can be found, for instance, in Ref.~\cite{Cruz:2020zkc}.
 
In the strong magnetic field regime, \(B_L \gg 1\),  the topological mass $m_B$ can be approximated as
\begin{eqnarray}
m_B^2&\simeq&-\frac{gB_L^2}{2^{\frac{1}{2}}\pi^{\frac{3}{2}}L^2z^{\frac{1}{2}}}e^{-z}\notag\\
&\simeq&-\frac{gB_L^{\frac{3}{2}}}{2\pi^{\frac{3}{2}}L^2}e^{-2B_L},
\label{TM_strong}
\end{eqnarray}
where the leading contribution is for $j=1$. This shows an exponential suppression with the field strength, as expected from the dominance of the lowest Landau level at high \(B\).

For the weak field regime, \(B_L \ll 1\), it is convenient to rewrite Eq.~\eqref{integral_B_topological} as
\begin{eqnarray}
\mathcal{I}_1(j, L, 0)  = 2B_L^2\int_{0}^{\infty} \frac{d\tau\,\tau^{-1}}{1 - e^{-2B_L^2\tau^2}} \,
e^{-(\mu_L^2 + B_L^2)\tau^2 - \frac{j^2}{\tau^2}}.
\label{integralTM_B0_zero10}
\end{eqnarray}
Using the expansion of Eq.~\eqref{series}, one obtains in dimensionless form, the expression
\begin{eqnarray}
4\pi^2L^2m_{\rm B}^2&\simeq& gz\sum_{j=1}^{\infty}(-1)^j\frac{K_1(zj)}{j} + 2gB_L^2\sum_{j=1}^{\infty}(-1)^jK_0(zj)\notag\\
 && + \frac{4gB_L^4}{3z}\sum_{j=1}^{\infty}(-1)^j jK_1(zj) -  \frac{16gB_L^8}{45z^3}\sum_{j=1}^{\infty}(-1)^j j^3K_3(zj).
\label{TMrenormalized_weak}
\end{eqnarray}
Expanding further, in analogy with the expansion steps used in Eqs.~\eqref{representation2}--\eqref{term31}, we find
\begin{eqnarray}
4\pi^2L^2m_{\rm B}^2&\simeq&2g\mu_L\sum_{j=1}^{\infty}(-1)^j\frac{K_1(2j\mu_L)}{j} + gB_L^2\ln(z)\notag\\
 && - \frac{gB_L^4}{3z^2}  -  \frac{gB_L^4}{3z} + \frac{116gB_L^8}{1215z^6} + \frac{232gB_L^8}{1215z^5}.
\label{TMrenormalized_weak2}
\end{eqnarray}

Now, setting \(\mu_L = 0\) in Eq.~\eqref{TMrenormalized_weak2}, for the massless case, we obtain
\begin{eqnarray}
4\pi^2L^2m_{\rm B}^2
\simeq -\frac{g\pi^2}{12} + gB_L^2\ln(2B_L) - \frac{1591gB_L^2}{19440} - \frac{1591gB_L^3}{9720}.
\label{TMrenormalized_weak3}
\end{eqnarray}
The leading constant term above is exactly twice the corresponding result obtained in Ref.~\cite{Cruz:2020zkc} for a real scalar field, as expected for a complex scalar field with the same boundary conditions.

The decomposition in Eq.~\eqref{TM} makes it explicit how the interplay between boundary conditions (through \(m_{\rm D}^2\)) and the magnetic field (through \(m_{\rm B}^2\)) controls the effective topological mass. In particular, the sign of \(m_{\rm T}^2\) can change depending on the competition between the Dirichlet term, which may induce spontaneous symmetry breaking in some regimes, and the magnetic term, which can either stabilize or destabilize the vacuum depending on the coupling \(g\) and the field strength. The region of the parameter space where the mass becomes negative has also been reported in other systems, where a vacuum stability analysis was performed~\cite{toms1980interacting, PhysRevD.107.125019}. Here, we leave such an analysis for future work, as previously mentioned. Regarding the limiting behaviors in Eqs.~\eqref{TM_strong}-\eqref{TMrenormalized_weak3}, they illustrate how the magnetic contribution is exponentially suppressed at strong fields but exhibits nontrivial polynomial and logarithmic corrections in the weak-field regime. These features are crucial for understanding phase transitions in confined quantum field systems subject to external magnetic fields.

\subsection{First-Order Coupling-Constant Corrections to the vacuum energy}
\label{secIVC}
%
The first-order coupling-constant corrections to the vacuum energy come from the two-loop level of the effective potential,
evaluated for the vacuum state $\Psi=0$, i.e.,
\begin{eqnarray}
V^{(2)}(\Psi) = V_{\lambda_{\psi}}^{(2)}(\Psi) + V_{\lambda_{\varphi}}^{(2)}(\Psi) + V_{g}^{(2)}(\Psi) + V_{2\lambda_{\varphi}}^{(2)}(\Psi),
\label{two-loop_total}
\end{eqnarray}
where $V_{\lambda_{\psi}}^{(2)}$
is the contribution from the self-interaction 
term $\frac{\lambda_{\psi}}{4!}\psi^4$ of the real field, 
$V_{\lambda_{\varphi}}^{(2)}$
is the contribution from the self-interaction of the complex field, namely $\frac{\lambda_{\varphi}}{4!}\varphi_1^4$
and $\frac{\lambda_{\varphi}}{4!}\varphi_2^4$,
$V_{g}^{(2)}$ is associated with the interaction between the
real and complex fields, $\frac{g}{2!}\varphi_1^2\psi^2$ and $\frac{g}{2!}\varphi_2^2\psi^2$,
and, finally, $V_{2\lambda_{\varphi}}^{(2)}$ is associated with the cross terms between the components of the
complex field, $\frac{\lambda_{\varphi}}{4!} 2\varphi_1^2\varphi_2^2$.

In terms of bubble vacuum diagrams, Eq.~\eqref{two-loop_total} can be rewritten as

\begin{eqnarray}
V^{(2)}(\Psi) = 
\underbrace{
\begin{tikzpicture}[baseline=-0.5ex]
  \draw[thick] (0,0) circle (0.4);
  \draw[thick] (0.8,0) circle (0.4);
  \fill (0.4,0) circle (2pt);
  \node at (-0.3,0.5) {\small \(\psi\)};
  \node at (1.0,0.55) {\small \(\psi\)};
\end{tikzpicture}}_{\psi\text{--}\psi \text{ loop}}
\quad +
\quad 2
\underbrace{
\begin{tikzpicture}[baseline=-0.5ex]
\draw[thick,dashed] (0,0) circle (0.4);
  \draw[thick,dashed] (0.8,0) circle (0.4);
  \fill (0.4,0) circle (2pt);
  \node at (-0.3,0.5) {\small \(\varphi\)};
  \node at (1.0,0.55) {\small \(\varphi\)};
\end{tikzpicture}}_{\varphi\text{--}\varphi \text{ loop}}
\quad+
\quad 2
\underbrace{
\begin{tikzpicture}[baseline=-0.5ex]
  \draw[thick] (0,0) circle (0.4);
  \draw[thick,dashed] (0.8,0) circle (0.4);
  \fill (0.4,0) circle (2pt);
  \node at (-0.4,0.5) {\small \(\psi\)};
  \node at (1.2,0.5) {\small \(\varphi\)};
\end{tikzpicture}}_{\psi\text{--}\varphi \text{ mixed loop}}
\quad+
\quad 2
\underbrace{
\begin{tikzpicture}[baseline=-0.5ex]
  \draw[thick] (0,0) circle (0.4);
  \draw[thick,dashed] (0.8,0) circle (0.4);
  \fill (0.4,0) circle (2pt);
  \node at (-0.4,0.5) {\small \(\varphi_1\)};
  \node at (1.2,0.5) {\small \(\varphi_2\)};
\end{tikzpicture}}_{\varphi_1\text{--}\varphi_2 \text{ mixed loop}}
\label{Fdiagrams}
\end{eqnarray}

We can use the zeta functions of the system to obtain the self-interaction term contributions by means of the expression
\begin{eqnarray}
V_{\rm C.C.}^{(2)}(0) = \frac{3\,\text{C.C.}}{4!} \left[ \frac{\zeta^{\rm R}(1)}{\Omega_4} \right]^2,
\label{zeta_general}
\end{eqnarray}
where C.C. denotes the coupling constant, and $\zeta^{\rm R}(1)$ is given by Eqs.~\eqref{zeta_D} and/or \eqref{zetaF_varphi}, depending on the specific term in Eq.~\eqref{two-loop_total} under consideration. In both cases, the terms proportional to $\Gamma(s-2)$ and $\Gamma(s-1)$ must be subtracted due to the divergent contributions at $s=1$, which correspond to the case of Minkowski spacetime without plates. Moreover, the second term on the r.h.s. of Eq.~\eqref{zeta_D} should also be subtracted, since it increases with the mass $m$.

The expression in Eq.~\eqref{zeta_general} can also be read directly from the Feynman diagrams for the self-interaction, represented by the first two terms on the r.h.s. of Eq.~\eqref{Fdiagrams}. The factor of three comes from the Wick theorem for field contractions, while the symmetry factor for this case is $\frac12$. Each loop corresponds to a zeta function \eqref{zeta_D} or \eqref{zetaF_varphi}, divided by $\Omega_4$, depending on the field type. In the corresponding diagrams, there is a single vertex contributing with a factor $\frac{\text{C.C.}}{4!}$.

For the contribution coming from $V_{\lambda_{\psi}}^{(2)}(\Psi)$, using Eqs.~\eqref{zeta_D} and \eqref{zeta_general}, we have 
\begin{eqnarray}
\frac{E_{\rm R}^{(2)}}{A} &=& L\,V_{\lambda_{\psi}}^{(2)}(0) \notag\\
&=& \frac{\lambda_{\psi} m^2}{128\pi^4 L} \left[ \sum_{n=1}^{\infty} \frac{K_1\!\left( 2n m L \right)}{n} \right]^2,
\label{lambda_psi_massive}
\end{eqnarray}
for the massive case, and for the massless case
\begin{eqnarray}
\frac{E_{\rm R}^{(2)}}{A} = \frac{\lambda_{\psi}}{18432\,L^3},
\label{lambda_psi_massless}
\end{eqnarray}
where we have used the Riemann zeta function $\zeta(2) = \frac{\pi^2}{6}$.

For the contribution coming from $V_{\lambda_{\varphi}}^{(2)}(\Psi)$, using Eqs.~\eqref{zetaF_varphi} and \eqref{zeta_general}, we obtain 
\begin{eqnarray}
\frac{E_{\rm R}^{(2)}}{A} = \frac{2\lambda_{\varphi}}{128\pi^4 L^3} 
\left\{ \sum_{j=1}^{\infty} (-1)^j \mathcal{I}_{1}(j, L, 0) \right\}^2,
\label{lambda_varphi_self}
\end{eqnarray}
where the factor of 2 multiplying $\lambda_{\varphi}$ accounts for the two components of the complex field, and $\mathcal{I}_{1}(j, L, 0)$ has been defined in Eq.~\eqref{integral_B_topological}. 

Let us now turn to the mixed terms. The first diagram representing mixed terms in Eq.~\eqref{Fdiagrams} indicates that the corresponding expression involves the product of the coupling constant $\frac{g}{2}$ (from a single vertex) by the renormalized zeta functions \eqref{zeta_D} and \eqref{zetaF_varphi} at $s=1$, divided by $\Omega_4^2$, with an extra factor of 2 accounting for the interaction of each component of $\varphi$ with $\psi$. In this case we have
\begin{eqnarray}
\frac{E_{\rm R}^{(2)}}{A} = \frac{2g\,m}{32\pi^4 L^2} 
\left[ \sum_{n=1}^{\infty} \frac{K_1\!\left( 2n m L \right)}{n} \right]
\left[ \sum_{j=1}^{\infty} (-1)^j \mathcal{I}_{1}(j, L, 0) \right].
\label{lambda_g_massive}
\end{eqnarray}

For the massless case, that is, $m=\mu=0$, we can make use of Eq.~\eqref{limit} and obtain
\begin{eqnarray}
\frac{E_{\rm R}^{(2)}}{A} = \frac{g}{192\pi^2 L^3} 
\left[ \sum_{j=1}^{\infty} (-1)^j \mathcal{I}_{1}(j, L, 0) \right],
\label{lambda_g_massless}
\end{eqnarray}
with $\mathcal{I}_{1}(j, L, 0)$ given by Eq.~\eqref{integral_B_topological} for $\mu=0$.

Finally, the mixed term corresponding to the last diagram in Eq.~\eqref{Fdiagrams} involves the product of the coupling constant $\frac{\lambda_{\varphi}}{4!}$ (from one vertex) by the square of the zeta function \eqref{zetaF_varphi} (one for each component of the complex field), with an additional factor of 2 accounting for both components. That is, 
\begin{eqnarray}
\frac{E_{\rm R}^{(2)}}{A} = \frac{2\lambda_{\varphi}}{384\pi^4 L^3} 
\left[ \sum_{j=1}^{\infty} (-1)^j \mathcal{I}_{1}(j, L, 0) \right]^2.
\label{lambda_varphi_mixed}
\end{eqnarray}

The asymptotic limits for the magnetic contribution in the above expressions follow the same qualitative behavior as in the topological mass case, where the function $\mathcal{I}_{1}(j, L, 0)$ was analyzed. In the strong-field regime, the expressions decay exponentially, and in the cases of Eqs.~\eqref{lambda_varphi_self} and \eqref{lambda_varphi_mixed} the decay is even faster. In the weak-field regime, the leading contribution for $\mathcal{I}_{1}(j, L, 0)$ is proportional to $B_L^2 \ln(z)$, which in Eqs.~\eqref{lambda_varphi_self} and \eqref{lambda_varphi_mixed} becomes $B_L^4 (\ln z)^2$, and is therefore even more suppressed.  

In summary, the expressions derived in this subsection constitute the complete set of first-order coupling-constant corrections to the vacuum energy at two-loop order, after removing the Minkowski–spacetime divergences and the mass-dependent terms that do not contribute to the renormalized results.
%

\section{Conclusions and Remarks}
\label{secV}

In this work, we investigated the interplay between a uniform magnetic field, boundary conditions, and scalar field interactions in the framework of quantum field theory in confined geometries. The model consists of a real scalar field satisfying Dirichlet boundary conditions, interacting via self- and cross-couplings with a gauge-coupled complex scalar field subjec to mixed boundary conditions between two perfectly reflecting parallel plates. Using the effective potential formalism and zeta-function regularization, we have obtained the renormalized effective potential up to first order in the coupling constants without imposing the vanishing of the magnetic field in the renormalization conditions, thereby preserving magnetic contributions while consistently removing divergences. 

We computed the vacuum energy per unit area of the plates, incorporating finite-size effects, magnetic contributions, and interaction corrections, and demonstrated the generation of a topological mass arising from the combined influence of boundary conditions and magnetic interactions. The results for the vacuum energy include both one-loop and two-loop (first-order in the couplings) corrections, showing that the magnetic field not only reorganizes the mode spectrum via Landau quantization but also qualitatively modifies the finite-size corrections. The asymptotic analysis revealed that in the strong-field regime the corrections to the vacuum energy are exponentially suppressed, whereas in the weak-field limit the behavior involves both polynomial and logarithmic terms, reflecting the subtle interplay between magnetic and boundary effects. These features may have implications for condensed-matter systems under strong magnetic fields, as well as for high-energy and cosmological scenarios where confined quantum fields interact with external background fields. 

The topological mass generated through boundary and magnetic effects displays an interesting dependence on the coupling constants and field strengths. The competition between contributions from the Dirichlet-confined field and the magnetically coupled field can lead to sign changes in the effective mass, suggesting possible vacuum instabilities or symmetry breaking phenomena. This warrants further investigation of the vacuum stability across different parameter regimes.

Our calculation of first-order coupling-constant corrections to the vacuum energy completes our investigation of quantum effects up to second order in perturbation theory. The systematic treatment developed here for coupled fields with different boundary conditions provides a framework that could be extended to more complex systems involving fermionic fields or non-Abelian gauge fields. Furthermore, future work could address finite-temperature regimes and explore the system in greater detail from a more realistic perspective, potentially uncovering additional physical phenomena arising from the relation between boundary conditions, magnetic fields, and interactions.

\acknowledgments

H.F.S.M. is partially supported by the Brazilian agency National Council for Scientific and Technological Development (CNPq) under Grant No. 308049/2023-3.


\end{document}